\documentclass[journal=jctcce,manuscript=article]{achemso}

\SectionNumbersOn

\usepackage[version=3]{mhchem} 
\usepackage{xcolor}
\usepackage{booktabs}

\author{Yu Jin}
\affiliation{Department of Chemistry, University of Chicago, Chicago, Illinois 60637, United States}
\author{Victor Wen-zhe Yu}
\affiliation{Materials Science Division, Argonne National Laboratory, Lemont, Illinois 60439, United States}
\author{Marco Govoni}
\email{mgovoni@unimore.it}
\affiliation{Department of Physics, Computer Science, and Mathematics, University of Modena and Reggio Emilia,
Modena, 41125, Italy}
\alsoaffiliation{Pritzker School of Molecular Engineering, University of Chicago, Chicago, Illinois 60637, United States}
\alsoaffiliation{Materials Science Division, Argonne National Laboratory, Lemont, Illinois 60439, United States}
\author{Andrew C Xu}
\affiliation{Pritzker School of Molecular Engineering, University of Chicago, Chicago, Illinois 60637, United States}
\author{Giulia Galli}
\email{gagalli@uchicago.edu}
\affiliation{Pritzker School of Molecular Engineering, University of Chicago, Chicago, Illinois 60637, United States}
\alsoaffiliation{Department of Chemistry, University of Chicago, Chicago, Illinois 60637, United States}
\alsoaffiliation{Materials Science Division, Argonne National Laboratory, Lemont, Illinois 60439, United States}

\title[An \textsf{achemso} demo]
  {Excited state properties of point defects in semiconductors and insulators investigated with time-dependent density functional theory}

\abbreviations{IR,NMR,UV}

\begin{document}

\begin{tocentry}

\includegraphics[width=8cm,height=4.5cm]{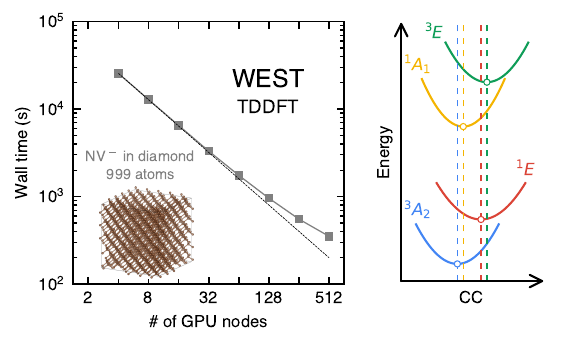}

\end{tocentry}

\begin{abstract}
We present a formulation of spin-conserving and spin-flip, hybrid time-dependent density functional theory (TDDFT), including the calculation of analytical forces, which allows for efficient calculations of excited state properties of solid-state systems with hundreds to thousands of atoms. We discuss an implementation on both GPU and CPU based architectures, along with several acceleration techniques. We then apply our formulation to the study of several point defects in semiconductors and insulators, specifically the negatively charged nitrogen-vacancy and neutral silicon-vacancy centers in diamond, the neutral divacancy center in 4H silicon carbide, and the neutral oxygen-vacancy center in magnesium oxide. Our results highlight the importance of taking into account structural relaxations in excited states, in order to interpret and predict optical absorption and emission mechanisms in spin-defects.    
\end{abstract}

\section{Introduction}
\label{sec:introduction}
Optically active point defects in semiconductors and insulators present a promising avenue for the development of quantum technologies~\cite{wolfowicz2021quantum}. Specific examples include the negatively charged nitrogen-vacancy center (NV$^-$) in diamond and the neutral divacancy center (VV$^0$) in 4H silicon carbide (SiC), both of which have been extensively studied~\cite{doherty2013nitrogen,gali2019ab,son2020developing}, for applications ranging from quantum sensing~\cite{schirhagl2014nitrogen,barry2020sensitivity} and quantum communication~\cite{childress2013diamond,christle2017isolated,wolfowicz2017optical,anderson2022five}, to potentially quantum computation~\cite{weber2010quantum,waldherr2014quantum}. These point defects may function as quantum bits (qubits), with an optical spin-polarization cycle for initialization and readout that involves both radiative and non-radiative transitions between many-body ground and excited states~\cite{thiering2018theory}. A comprehensive understanding of the physics underlying these transitions is essential to interpret experiments and to formulate general guidelines for the use of point defects in quantum technologies~\cite{manson2006nitrogen,maze2011properties,doherty2011negatively}.

First-principles methods have significantly contributed to the understanding of the electronic structure of the ground and excited states of point defects in a variety of semiconductors and insulators~\cite{freysoldt2014first,alkauskas2016tutorial,dreyer2018first,gali2023recent}. Many first-principles studies on the electronic excited states of point defects have used the constrained occupation density functional theory (CDFT) method, often called $\Delta$SCF, where the occupations of localized defect orbitals are adjusted to represent a specific excited state~\cite{gali2009theory}. Similar to ground state DFT calculations, the $\Delta$SCF approach has the advantage of a relatively low computational cost; in addition, it allows for the straightforward calculation of analytical nuclear forces and thus for geometry optimizations, vibrational mode calculations in excited states, and for the investigation of vibrationally resolved optical spectra~\cite{alkauskas2014first,razinkovas2021vibrational,jin2021pl}. However, the $\Delta$SCF approach is only applicable to excited states that are well approximated by a single Slater determinant, and it is prone to numerical convergence issues in the case of (near-)degenerate excited states. Other, potentially more accurate approaches are based on quantum embedding theories~\cite{sun2016quantum,vorwerk2022quantum}, e.g. the quantum defect embedding theory (QDET)~\cite{ma2020quantum,ma2021qdet,sheng2022edc}, and the density matrix embedding theory (DMET)~\cite{knizia2012density,knizia2013density,pham2019periodic}, where an active space representing the manifold of defect states is identified, and such states are treated at a higher level of theory than the electronic structure of the rest of the host crystal. For example, one may use Full-Configuration Interaction (FCI) for an accurate description of multi-configuration excited states of point defects in semiconductor and oxides~\cite{ma2020first,mitra2021excited,bhang2021first,huang2022simulating,huang2023quantum,haldar2023local,verma2023optical}, with the rest of the crystal treated at the DFT or many-body perturbation theory ($GW$) level of theory. However, embedding methods have so far been limited to single-point calculations at fixed atomic geometries, and a formalism to carry out geometry optimizations in electronic excited states is not yet available.

Linear-response time-dependent density functional theory (LR-TDDFT) is a computational method frequently used to study excited state and optical properties of molecules and solids~\cite{casida1995tddft,casida1996time}. In particular, spin-conserving and spin-flip TDDFT allows for the description of excited states as linear combinations of single excitations with different spin-multiplicities~\cite{shao2003spin,wang2004time,wang2005performance,seth2011time,li2012theoretical,bernard2012general,casanova2020spin}. In several cases, it has been shown that the accuracy of the TDDFT method using hybrid functionals is comparable to that of higher-level electronic structure methods, such as the solution of the Green’s function based $GW$ plus Bethe--Salpeter Equation ($GW$/BSE) and equation-of-motion coupled cluster (EOM-CC) methods~\cite{yang2015simple,sun2020low,caricato2011oscillator,maier2016validation}. The computational workload of TDDFT calculations scales as $\mathcal{O}(N_{\text{occ}}^2 N_g\log(N_g))$ when using hybrid functionals and a plane-wave basis set, where $N_{\text{occ}}$ is the number of occupied Kohn-Sham orbitals of the system, and $N_g$ is the number of plane-waves; such scaling is comparable to that of ground state DFT calculations. Further, analytical nuclear forces on nuclei can be computed within TDDFT, allowing for geometry optimizations and the calculation of nuclear vibrations in electronic excited states~\cite{hutter2003excited}.

TDDFT has been used to study excited states of point defects in diamond using cluster models and atomic-centered basis sets~\cite{gali2011time,petrone2016quantum,petrone2018electronic,reimers2020photoluminescence,karim2020ab,karim2021bright,beck2023electronic}. However, depending on the size and the system, cluster models may exhibit quantum confinement and their electronic structure may also be impacted by the chosen surface termination. In addition, the optimized geometries in the ground and excited states may differ from the corresponding ones in the solid state. Although TDDFT and analytical nuclear forces have been implemented with a plane-wave pseudopotential method for calculations of solid-state systems with periodic boundary conditions~\cite{hutter2003excited,zhang2015subspace,arhangelskis2018time,zhang2020linear,liu2023efficient}, the direct use of TDDFT for modeling point defects in calculations with periodic boundary conditions is still rare.

In this work, we present an efficient numerical implementation of spin-conserving and spin-flip TDDFT, including the evaluation of analytical nuclear forces, in the open-source code WEST~\cite{govoni2015west,yu2022west}, which is based on the plane-wave pseudopotential method. Hereafter, we refer to our implementation of TDDFT as WEST-TDDFT. We adopt several numerical approximations to accelerate TDDFT calculations with hybrid functionals in WEST-TDDFT, including the adaptively compressed exchange (ACE) operator~\cite{lin2016adaptively}, the use of the near-sightedness principle~\cite{nguyen2019finite}, and of the inexact Krylov subspace approach~\cite{simoncini2003theory,vandeneshof2004inexact}. We show that the errors introduced by these approximations can be systematically controlled and that high-accuracy results may be obtained. We show that all together, these approximations lead to about an order of magnitude speed-up in the computation of excited state energies and analytical nuclear forces for the defective systems benchmarked in this work. We also show that with a multilevel parallelization scheme~\cite{yu2022west}, WEST-TDDFT shows strong scaling up to hundreds of CPU and GPU nodes, enabling the study of point defects for systems containing hundreds to thousands of atoms.

To demonstrate the capabilities of WEST-TDDFT, we investigated the excited states and optical properties of several point defects in solids, including the NV$^-$ and the neutrally charged silicon-vacancy center (SiV$^0$) in diamond, the VV$^0$ in 4H-SiC, and the neutrally charged oxygen vacancy center (V$_{\text{O}}^0$) in magnesium oxide (MgO). We used the dielectric dependent hybrid (DDH) functional in TDDFT calculations, where the coefficient of the Hartree-Fock exact exchange was set to the inverse of the high frequency dielectric constant of the host material. DDH has been shown to improve the description of the electronic structure of a broad range of systems~\cite{skone2014ddh,skone2016ddh}, including systems with impurities~\cite{seo2017designing,gaiduk2016photoelectron,gaiduk2018first,gerosa2018role,pham2017electronic,jin2021pl}. Hence, using the DDH functional in TDDFT calculations is expected to lead to a more accurate description of excitonic effects than using semi-local functionals, due to the inclusion of screening effects in the Coulomb interaction between the electron and the hole in the excited state~\cite{sun2020low,tal2020accurate,dong2021machine}. Our results obtained with TDDFT using the DDH functional offer insights into the mechanisms underlying the optical absorption and emission processes in these defects.

The rest of the paper is organized as follows. Section~\ref{sec:method} introduces the methodology. Section~\ref{sec:numericalapproximations} discusses the numerical approximations adopted in WEST-TDDFT to accelerate the calculations for hybrid functionals. In section~\ref{sec:performance}, we present the parallelization and scaling tests of WEST-TDDFT. In section~\ref{sec:results}, we give examples of applying WEST-TDDFT to study the excited state properties of prototypical point defects in solids. We close the paper with a summary of all the results and an outlook on future work.


\section{Method}
\label{sec:method}

We describe below the formalism adopted to compute vertical excitation energies (subsection 2.1) and nuclear forces (subsection 2.2) with TDDFT and with spin-flip TDDFT (subsection 2.3). Since our goal is to study excited states of large systems, such as point defects in solids in the dilute limit, we performed TDDFT calculations using only the $\Gamma$ point to sample the Brillouin zone.

\subsection{Vertical excitation energies} \label{subsec:lrtddft}
Within LR-TDDFT the vertical excitation energy (VEE) $\omega_s$ from the ground to the $s$-th excited state,  can be obtained by solving the following eigenvalue problem~\cite{casida1995tddft,casida1996time,gross2005density}:
\begin{equation}
    \left(
    \begin{matrix}
    \mathcal{D} + \mathcal{K}^{1e} - \mathcal{K}^{1d} & \mathcal{K}^{2e} - \mathcal{K}^{2d} \\
    \mathcal{K}^{2e} - \mathcal{K}^{2d} & \mathcal{D} + \mathcal{K}^{1e} - \mathcal{K}^{1d} \\
    \end{matrix}
    \right) \left(
    \begin{matrix}
    \mathcal{A}_s \\
    \mathcal{B}_s
    \end{matrix}
    \right) = \omega_s \left(
    \begin{matrix}
    \mathcal{I} & 0 \\
    0 & -\mathcal{I}
    \end{matrix}
    \right) \left(
    \begin{matrix}
    \mathcal{A}_s \\
    \mathcal{B}_s
    \end{matrix}
    \right),
    \label{eq:TDDFT_VEE}
\end{equation}
where $\mathcal{A}_s = \left\{|a_{s,v\sigma}\rangle: v=1,\dots, N_{\text{occ},\sigma}; \sigma = \uparrow, \downarrow \right\}$ and $\mathcal{B}_s = \left\{|b_{s,v\sigma}\rangle: v=1,\dots, N_{\text{occ},\sigma}; \sigma = \uparrow, \downarrow \right\}$ are two sets of orbitals that enter the definition of the linear change of the density matrix in the $\sigma$ spin channel with respect to the ground state density matrix, due to the $s$-th neutral excitation:
\begin{equation}
    \Delta \rho_{s, \sigma}  = \sum_{v=1}^{N_{\text{occ},\sigma}} \left(  |a_{s,v\sigma}\rangle \langle \varphi_{v\sigma}| + |\varphi_{v\sigma}\rangle \langle b_{s,v\sigma}| \right),
    \label{eq:deltarho}
\end{equation}
Here $N_{\text{occ},\sigma}$ is the number of occupied states in the $\sigma$ spin channel, and $|\varphi_{v\sigma}\rangle$ are the Kohn-Sham (KS) wavefunctions of the ground state. 

In this section, we consider neutral excitations, i.e., excitations that do not change the total number of electrons, and $N_{\text{occ},\sigma}$ is individually conserved for each spin channel. Neutral excitations that flip the spin will be discussed in section~\ref{subsec:spin-flip}.

The operators on the left-hand side (LHS) of Eq.~\eqref{eq:TDDFT_VEE} are defined as:
\begin{equation}
    \mathcal{DA}_s = \left\{\mathcal{P}^c_{\sigma} \left( H_{\sigma}^{\text{KS}} - \varepsilon_{v\sigma}\right)|a_{s,v\sigma} \rangle: v=1,\dots, N_{\text{occ},\sigma}; \sigma = \uparrow, \downarrow \right\},
    \label{eq:dterm}
\end{equation}
\begin{equation}
    \begin{aligned}
        &\mathcal{K}^{1e}\mathcal{A}_s = \\ &\left\{ \int \mathrm{d} \mathbf{r}^{\prime} \mathcal{P}^c_{\sigma}(\mathbf{r,r'}) \varphi_{v\sigma}(\mathbf{r'}) \sum_{\sigma'} \int \mathrm{d} \mathbf{r''} f_{\text{Hxc},\sigma\sigma'}^{\text{loc}}(\mathbf{r',r''}) \sum_{v'=1}^{N_{\text{occ},\sigma'}} \varphi_{v'\sigma'}^{\ast} (\mathbf{r''}) a_{s,v'\sigma'} (\mathbf{r''}) : v = 1, \dots, N_{\text{occ},\sigma}; \sigma = \uparrow, \downarrow \right\},\\
    \end{aligned}
    \label{eq:k1eterm}
\end{equation}
\begin{equation}
    \begin{aligned}
    &\mathcal{K}^{2e}\mathcal{A}_s = \\&
    \left\{ \int \mathrm{d} \mathbf{r}^{\prime} \mathcal{P}^c_{\sigma}(\mathbf{r,r'}) \varphi_{v\sigma}(\mathbf{r'}) \sum_{\sigma'} \int \mathrm{d} \mathbf{r''} f_{\text{Hxc},\sigma\sigma'}^{\text{loc}}(\mathbf{r',r''}) \sum_{v'=1}^{N_{\text{occ},\sigma'}} a_{s,v'\sigma'}^{\ast} (\mathbf{r''}) \varphi_{v'\sigma'} (\mathbf{r''}) : v = 1, \dots, N_{\text{occ},\sigma}; \sigma = \uparrow, \downarrow \right\},\\
    \end{aligned}
    \label{eq:k2eterm}
\end{equation}
\begin{equation}
    \label{eq:k1d}
    \mathcal{K}^{1d}\mathcal{A}_s = \left\{ \alpha_{\text{EXX}} \int \mathrm{d} \mathbf{r'} \mathcal{P}^c_{\sigma}(\mathbf{r,r'}) \sum_{v'=1}^{N_{\text{occ},\sigma}} a_{s,v'\sigma} (\mathbf{r'}) \int \mathrm{d} \mathbf{r''} v_c(\mathbf{r',r''}) \varphi^{\ast}_{v'\sigma}(\mathbf{r''}) \varphi_{v\sigma}(\mathbf{r''}): v = 1, \dots, N_{\text{occ},\sigma}; \sigma = \uparrow, \downarrow \right\},
\end{equation}
\begin{equation}
    \mathcal{K}^{2d}\mathcal{A}_s = \left\{ \alpha_{\text{EXX}} \int \mathrm{d} \mathbf{r'} \mathcal{P}^c_{\sigma}(\mathbf{r,r'}) \sum_{v'=1}^{N_{\text{occ},\sigma}} \varphi_{v'\sigma} (\mathbf{r'}) \int \mathrm{d} \mathbf{r''} v_c(\mathbf{r',r''}) a_{s,v'\sigma}^{\ast}(\mathbf{r''}) \varphi_{v\sigma}(\mathbf{r''}): v = 1, \dots, N_{\text{occ},\sigma}; \sigma = \uparrow, \downarrow \right\},
    \label{eq:k2dterm}
\end{equation}
where $H_{\sigma}^{\text{KS}}$ is the ground state KS Hamiltonian written for the $\sigma$ spin channel, and $f_{\text{Hxc},\sigma\sigma'}^{\text{loc}} (\mathbf{r,r'}) = v_c(\mathbf{r,r'}) + f_{\text{xc},\sigma\sigma'}^{\text{loc}} (\mathbf{r,r'})$ is the sum of the bare Coulomb potential, $v_c$, and the local part of the time-independent exchange-correlation kernel
\begin{equation}
    f_{\text{xc},\sigma\sigma'}^{\text{loc}} (\mathbf{r,r'}) = \left. \dfrac{\delta V_{\text{xc},\sigma}^{\text{loc}}(\mathbf{r})}{\delta \rho_{\sigma'} (\mathbf{r'})} \right |_{\left(\rho^0, \nabla\rho^0\right)}\,.
\end{equation}
$\rho_{\sigma}(\mathbf{r})$ is the electron density of the spin channel $\sigma$, $\mathcal{P}^c_{\sigma}$ is the projection operator onto the unoccupied Kohn-Sham orbitals with $\sigma$ spin, i.e., $\mathcal{P}^c_{\sigma}=1-\sum_{v=1}^{N_{\text{occ},\sigma}}|\varphi_{v\sigma}\rangle\langle\varphi_{v\sigma}|$. The parameter $\alpha_{\text{EXX}}$ is the fraction of the Hartree-Fock exchange included in the definition of the hybrid functional. For semilocal functionals $\alpha_{\text{EXX}}=0$ and hence the $\mathcal{K}^{1d}$ and $\mathcal{K}^{2d}$ operators are zero by definition. 

Under the Tamm--Dancoff approximation (TDA)~\cite{Gross1990}, which is equivalent to the Configuration Interaction Singles (CIS) method~\cite{hirata1999configuration} used in quantum chemistry to compute VEEs, the $\mathcal{K}^{2e}$ and $\mathcal{K}^{2d}$ terms in Eq.~\eqref{eq:TDDFT_VEE} are neglected, yielding $\mathcal{B}_s=0$, and one solves the following eigenvalue problem:
\begin{equation}
    \left( \mathcal{D} + \mathcal{K}^{1e} - \mathcal{K}^{1d} \right) \mathcal{A}_s = \omega_s \mathcal{A}_s\,.
    \label{eq:TDDFT_VEXX_TDA}
\end{equation}
%


\subsection{Excited state nuclear forces} \label{subsec:forces}
To derive excited states nuclear forces, we use the extended Lagrangian formalism and the $Z$-vector method of Handy–Schaefer to compute gradients~\cite{handy1984evaluation}. Such formulation allows us to compute gradients of the excited potential energy surfaces described by Eq.~\eqref{eq:TDDFT_VEXX_TDA}, while satisfying all orthonormality constraints of the KS and $|a_{s,v\sigma}\rangle$ orbitals. The implementation of nuclear forces carried out in our work is similar to that of Hutter~\cite{hutter2003excited}, although we generalize it in section~\ref{subsec:spin-flip} to spin-flip TDDFT.

To obtain the nuclear force for the $I$-th atom along the $\alpha$-th Cartesian direction of a system in the $s$-th excited state, we need to evaluate the following total derivatives:
\begin{equation}
F_{s,I\alpha} = - \left(\frac{d V^{nn}}{d R_{I\alpha}} + \frac{d E^{\text{KS}}}{d R_{I\alpha}} + \frac{d \omega_s}{d R_{I\alpha}} \right),
\label{eq:nuclearforce}
\end{equation}
where $V^{nn}$ is the nuclear-nuclear electrostatic potential energy, $E^{\text{KS}}$ is the KS ground state total energy, and $\omega_s= \mathcal{A}_s^\dagger \left( \mathcal{D} + \mathcal{K}^{1e} - \mathcal{K}^{1d} \right) \mathcal{A}_s$ is the VEE of the $s$-th excited state obtained by solving the eigenvalue problem of Eq.~\eqref{eq:ltda} within the TDA. Without any loss of generality, we simplify the notation in Eq.~\eqref{eq:nuclearforce} and write all terms as total derivatives with respect to an external parameter $\tau$, and we define an extended Lagrangian, $\mathcal{L}_{\text{ex}}$ (the dependence on the excited state index $s$ is dropped for clarity)
\begin{equation}
F_{\tau} = - \left(\frac{d V^{nn}}{d \tau} + \frac{d \mathcal{L}_{\text{ex}}}{d \tau}\right),
\end{equation}
where 
\begin{equation}
\mathcal{L}_{\text{ex}} = \mathcal{L}_{\text{KS}} + \mathcal{L}_{\text{TDA}} + \mathcal{L}_{Z}
\end{equation}
and we have defined 
\begin{equation}
\label{eq:lks}
\mathcal{L}_{\text{KS}} = E^{\text{KS}} - \sum_{v \geq v^{{\prime}},\sigma} \Gamma_{vv^{\prime}\sigma}\left( \langle \varphi_{v\sigma} | \varphi_{v^{\prime}\sigma} \rangle - \delta_{vv^{\prime}} \right),
\end{equation}
\begin{equation}
\label{eq:ltda}
    \mathcal{L}_{\text{TDA}} =
    \sum_{vv^{\prime}\sigma\sigma^{\prime}} \langle a_{v\sigma} | (\mathcal{D} + \mathcal{K}^{1e} - \mathcal{K}^{1d})_{vv^{\prime}\sigma\sigma^{\prime}} | a_{v^{\prime}\sigma^{\prime}} \rangle - \omega \left( \sum_{v\sigma} \langle a_{v\sigma} | a_{v\sigma} \rangle - 1 \right),
\end{equation}
\begin{equation}
\label{eq:lz}
    \mathcal{L}_{Z} = \sum_{v\sigma} \langle Z_{v\sigma} | \left( H_{\sigma}^{\text{KS}} | \varphi_{v\sigma} \rangle - \sum_{v^{\prime}} \Lambda_{vv^{\prime}\sigma} |\varphi_{v^{\prime}\sigma} \rangle \right) + c.c.
\end{equation}
By requiring that $\mathcal{L}_{\text{ex}}$ is stationary with respect to $\langle{\varphi_{v\sigma}}|$, $\langle{a_{v\sigma}}|$, $\langle{Z_{v\sigma}}|$, $\omega$, $\Gamma_{vv^\prime\sigma}$ and $\Lambda_{vv^\prime\sigma}$ we obtain the following result:
\begin{equation}
\frac{d\mathcal{L}_{\text{ex}}}{d\tau} = \frac{\partial E^{\text{KS}}}{\partial\tau} +  
\int \mathrm{d} \mathbf{r} \dfrac{\partial V_{\text{ext}}(\mathbf{r})}{\partial \tau} \sum_{\sigma} \left[ \Delta \rho^{(x)}_{\sigma} (\mathbf{r}) +  \Delta \rho_{\sigma}^{(Z)} (\mathbf{r})\right].
\label{eq:Lexresult}
\end{equation}
In Eq.~\eqref{eq:Lexresult}, $\frac{\partial E^{\text{KS}}}{\partial \tau}$ is the partial derivative of the KS-DFT total energy with respect to nuclear coordinates obtained from the Hellmann-Feynman theorem, $\frac{\partial V_{\text{ext}} (\mathbf{r})}{\partial \tau}$ is the derivative of the external potential (described using pseudopotentials in our formulation), and we have defined
\begin{equation}
    \Delta \rho_{\sigma}^{(x)}(\mathbf{r}) = \sum_{v=1}^{N_{\text{occ},\sigma}} |a_{v\sigma}(\mathbf{r})|^2 - \sum_{v=1}^{N_{\text{occ},\sigma}} \sum_{v'=1}^{N_{\text{occ},\sigma}} \varphi^{\ast}_{v\sigma}(\mathbf{r}) \varphi_{v'\sigma}(\mathbf{r}) \int \mathrm{d} \mathbf{r'}  a^{\ast}_{v\sigma}(\mathbf{r'}) a_{v'\sigma}(\mathbf{r'}),
\end{equation}
and
\begin{equation}
    \Delta \rho_{\sigma}^{(Z)} (\mathbf{r}) = \sum_{v=1}^{N_{\text{occ},\sigma}} \left[Z_{v\sigma}^{\ast}(\mathbf{r}) \varphi_{v\sigma} (\mathbf{r}) + \varphi_{v\sigma}^{\ast} (\mathbf{r}) Z_{v\sigma} (\mathbf{r}) \right].
\end{equation}
Here $\mathcal{Z} = \left\{ |Z_{v\sigma}\rangle: v = 1, \dots, N_{\text{occ},\sigma}; \sigma = \uparrow, \downarrow \right\}$ are vectors obtained by solving the so-called Handy-Schaefer $Z$-vector equation (obtained by imposing that $\mathcal{L}_{\text{ex}}$ is stationary with respect to $\langle\varphi_{v\sigma}|$); the equation reads:
\begin{equation}
\label{eq:z-v-eq}
    \left( \mathcal{D} + \mathcal{K}^{1e} - \mathcal{K}^{1d} + \mathcal{K}^{2e} - \mathcal{K}^{2d} \right) \mathcal{Z} = \mathcal{U}.
\end{equation}
We have defined $\mathcal{U} = \left\{ - \mathcal{P}^c_{\sigma} |u_{v\sigma} \rangle: v = 1, \cdots, N_{\text{occ},\sigma} ; \sigma = \uparrow, \downarrow \right\}$, with
\begin{equation}
u_{v\sigma} (\mathbf{r}) = \dfrac{\delta \left[ \mathcal{A}^{\dagger} \left( \mathcal{D} + \mathcal{K}^{1e} - \mathcal{K}^{1d} \right) \mathcal{A}\right]}{\delta \varphi^{\ast}_{v\sigma}(\mathbf{r})} \,.
\end{equation}
A formal expression for $u_{v\sigma} (\mathbf{r})$ is reported in section~S1 of the Supporting Information (SI). 

\subsection{Vertical excitation energies and excited state nuclear forces within spin-flip TDDFT}
\label{subsec:spin-flip}
Here we extend the LR-TDDFT formalism derived in section~\ref{subsec:lrtddft} and~\ref{subsec:forces} to include neutral excitations involving a  flip of the spin state, i.e., excitations that bring an electron from the spin channel $\sigma$ to the opposite spin channel $\tilde{\sigma}$, without changing the total number of electrons~\cite{shao2003spin,wang2004time,wang2005performance,seth2011time,li2012theoretical,bernard2012general,casanova2020spin}.
In the presence of spin-flip (SF) excitations, the LR-TDDFT eigenvalue equation within the TDA, Eq.~\eqref{eq:TDDFT_VEXX_TDA}, reads
\begin{equation}
    \left( \mathcal{D}^{\text{SF}} + \mathcal{K}^{1e,\text{SF}} - \mathcal{K}^{1d, \text{SF}} \right) \mathcal{A}_s^{\text{SF}} = \omega_s \mathcal{A}_s^{\text{SF}}\,.
    \label{eq:TDDFT_VEXX_TDA_SF}
\end{equation}
The eigenvalue equation without the TDA can be found in section~S1 of the SI. The variation of the electron density can be written as
\begin{equation}
    \Delta \rho_{s,\sigma}^{\text{SF}} (\mathbf{r}) = \sum_{v=1}^{N_{\text{occ},\widetilde{\sigma}}} \varphi_{v\widetilde{\sigma}}^{\ast} (\mathbf{r}) a_{s,v\sigma}^{\text{SF}} (\mathbf{r}).
\end{equation}
The operators on the LHS of Eq.~\eqref{eq:TDDFT_VEXX_TDA_SF} are defined as
\begin{equation}
    \mathcal{D}^{\text{SF}} \mathcal{A}_s^{\text{SF}} = \left\{ \mathcal{P}_{\sigma}^c \left(H_{\sigma}^{\text{KS}} - \varepsilon_{v\widetilde{\sigma}} \right) | a_{s,v\sigma}^{\text{SF}} \rangle : v = 1, \dots, N_{\text{occ},\widetilde{\sigma}}; \sigma = \uparrow, \downarrow \right\},
\end{equation}
\begin{equation}
\begin{aligned}
    &\mathcal{K}^{1e,\text{SF}}\mathcal{A}_s^{\text{SF}} \\
    &= \left\{ \int \mathrm{d} \mathbf{r}^{\prime} \mathcal{P}^c_{\sigma}(\mathbf{r,r'}) \varphi_{v\widetilde{\sigma}}(\mathbf{r'}) \int \mathrm{d} \mathbf{r''} f_{\text{xc}}^{\text{loc,SF}}(\mathbf{r',r''}) \sum_{v'=1}^{N_{\text{occ},\widetilde{\sigma}}} \varphi_{v'\widetilde{\sigma}}^{\ast} (\mathbf{r}'') a_{s,v'\sigma}^{\text{SF}} (\mathbf{r}'') : v = 1, \dots, N_{\text{occ},\widetilde{\sigma}}; \sigma = \uparrow, \downarrow \right\},\\
\end{aligned}
\end{equation}
\begin{equation}
\begin{aligned}
    &\mathcal{K}^{1d,\text{SF}}\mathcal{A}_s^{\text{SF}} \\
    &= \bigg\{ \alpha_{\text{EXX}} \int \mathrm{d} \mathbf{r'} \mathcal{P}^c_{\sigma}(\mathbf{r,r'})\sum_{v'=1}^{N_{\text{occ},\widetilde{\sigma}}} a_{s,v'\sigma}^{\text{SF}} (\mathbf{r'}) \int \mathrm{d} \mathbf{r''} v_c(\mathbf{r',r''}) \varphi^{\ast}_{v'\widetilde{\sigma}}(\mathbf{r''}) \varphi_{v\widetilde{\sigma}}(\mathbf{r''}):
    v = 1, \dots, N_{\text{occ},\widetilde{\sigma}}; \sigma = \uparrow, \downarrow \bigg\}.\\
\end{aligned}
\end{equation}
The spin-flip exchange-correlation kernel is defined as
\begin{equation}
    f_{\text{xc}}^{\text{loc,SF}} (\mathbf{r,r'}) = \left.\dfrac{V_{\text{xc},\sigma}^{\text{loc}}(\mathbf{r}) - V_{\text{xc},\widetilde{\sigma}}^{\text{loc}}(\mathbf{r})}{\rho_{\sigma}(\mathbf{r}) - \rho_{\widetilde{\sigma}}(\mathbf{r}) } \right|_{\left( \rho^0,\nabla\rho^0\right)} \delta (\mathbf{r,r'}), \quad \sigma \neq \widetilde{\sigma}.
\end{equation}
Nuclear forces are evaluated using Eq.~\eqref{eq:nuclearforce}, but using the following expression for $\Delta \rho_{\sigma}^{(x)}$
\begin{equation}
    \Delta \rho_{\sigma}^{(x)}(\mathbf{r}) = \sum_{v=1}^{N_{\text{occ},\widetilde{\sigma}}} | a_{v\sigma}^{\text{SF}} (\mathbf{r})|^2 - \sum_{v=1}^{N_{\text{occ},\sigma}} \sum_{v'=1}^{N_{\text{occ},\sigma}} \varphi^{\ast}_{v\sigma}(\mathbf{r}) \varphi_{v'\sigma}(\mathbf{r}) \int \mathrm{d} \mathbf{r'}  a^{\text{SF}\ast}_{v\widetilde{\sigma}}(\mathbf{r'}) a_{v'\widetilde{\sigma}}^{\text{SF}} (\mathbf{r'}).
\end{equation}
The $Z$-vectors are obtained by solving Eq.~\eqref{eq:z-v-eq}, and using the definition for $u_{v\sigma}$ derived in the presence of spin-flip transitions (see section~S1 of the SI). 

\section{Numerical approximations to compute vertical excitation energies and excited state nuclear forces}
\label{sec:numericalapproximations}

In this section we discuss numerical approximations adopted in WEST-TDDFT for the calculation of VEEs with Eq.~\eqref{eq:TDDFT_VEXX_TDA} or Eq.~\eqref{eq:TDDFT_VEXX_TDA_SF} for spin-conserving or spin-flip excitations, respectively, and excited state nuclear forces with Eq.~\eqref{eq:nuclearforce}. In particular, we focus on the most computationally expensive terms in the case of hybrid functionals, i.e., the $\mathcal{D}$, the $\mathcal{K}^{1d}$, and the $\mathcal{K}^{2d}$ terms, as defined in Eqs.~\eqref{eq:dterm}, ~\eqref{eq:k1d}, and ~\eqref{eq:k2dterm}, respectively. In the following, we present the impact of the approximations and our verification tests performed for the first triplet excited state of the NV$^-$ in diamond, using a conventional $(3\times3\times3)$ supercell containing 215 atoms, and the dielectric dependent hybrid (DDH) functional~\cite{skone2014ddh,skone2016ddh}, with a plane-wave energy cutoff set to 60 Ry. TDDFT calculations were performed under the TDA.

\subsection{Approximating the $\mathcal{D}$ term}
When a hybrid functional is used, the Kohn-Sham Hamiltonian $H_\sigma^{\text{KS}}$ that appears in the $\mathcal{D}$ term of Eq.~\eqref{eq:dterm} contains the exact exchange operator, $V_{\text{X},\sigma}$, whose application to an arbitrary function, $\psi_\sigma(\mathbf{r'})$, is defined as
\begin{equation}
    \int V_{\text{X},\sigma} (\mathbf{r,r}') \psi_\sigma (\mathbf{r}') \mathrm{d} \mathbf{r'} = - \sum_{v}^{N_{\text{occ},\sigma}} \varphi_{v\sigma} (\mathbf{r}) \int \dfrac{\varphi^\ast_{v\sigma} (\mathbf{r'}) \psi_{\sigma} (\mathbf{r'})}{|\mathbf{r - r'}|} \mathrm{d} \mathbf{r'}\,.
    \label{eq:vexx}
\end{equation}
To reduce the computational cost associated with the $\mathcal{D}$ term, we approximated $V_{\text{X},\sigma}$ using the ACE operator~\cite{lin2016adaptively}, defined as
\begin{equation}
    V_{\text{X},\sigma}^{\text{ACE}} (\mathbf{r,r'}) = - \sum_{k=1}^{N_{\text{ACE},\sigma}} \xi_{k\sigma} (\mathbf{r}) \xi^\ast_{k\sigma} (\mathbf{r'}),
    \label{eq:ace}
\end{equation}
where $N_{\text{ACE},\sigma}$ is the total number of KS orbitals included in the summation that defines the operator. The ACE orbitals $\xi_{k\sigma}$ are obtained as:
\begin{equation}
    \xi_{k\sigma} (\mathbf{r}) = \sum_{i=1}^{N_{\text{ACE},\sigma}} W_{i\sigma} (\mathbf{r}) \left( L^{-T}_\sigma \right)_{ik},
\end{equation}
where $W_{i\sigma} (\mathbf{r}) = \int V_{\text{X},\sigma}(\mathbf{r,r'})\varphi_{i\sigma}(\mathbf{r'}) \mathrm{d} \mathbf{r'}$ results from the application of $V_{\text{X},\sigma}$ to KS orbitals in the range of $[1, N_{\text{ACE},\sigma}]$, and $\mathbf{L}_{\sigma}$ is the lower triangular matrix obtained from the Cholesky factorization $\mathbf{M}_\sigma = -\mathbf{L}_\sigma \mathbf{L}_\sigma^T$, where $\mathbf{M}_{\sigma}$ is a $N_{\text{ACE},\sigma} \times N_{\text{ACE},\sigma}$ matrix with $M_{ij\sigma} = \int \varphi^\ast_{i\sigma} (\mathbf{r}) W_{j\sigma} (\mathbf{r}) \mathrm{d} \mathbf{r}$.

In this work we use the ACE operator, $V_{\text{X},\sigma}^{\text{ACE}}$, to approximate the exact exchange operator, $V_{\text{X},\sigma}$, in the $\mathcal{D}$ term of Eq.~\eqref{eq:dterm}. Since $V_{\text{X},\sigma}$ is to be applied to functions $a_{v\sigma}(\mathbf{r}')$ that are orthogonal to the occupied KS orbitals, unoccupied KS orbitals in the range of $[ N_{\text{occ},\sigma} + 1, N_{\text{ACE},\sigma}]$ are included in the construction of the ACE operator, in addition to the occupied KS orbitals in the range of $[1, N_{\text{occ},\sigma}]$.

\begin{figure}
    \centering
    \includegraphics[width=14cm]{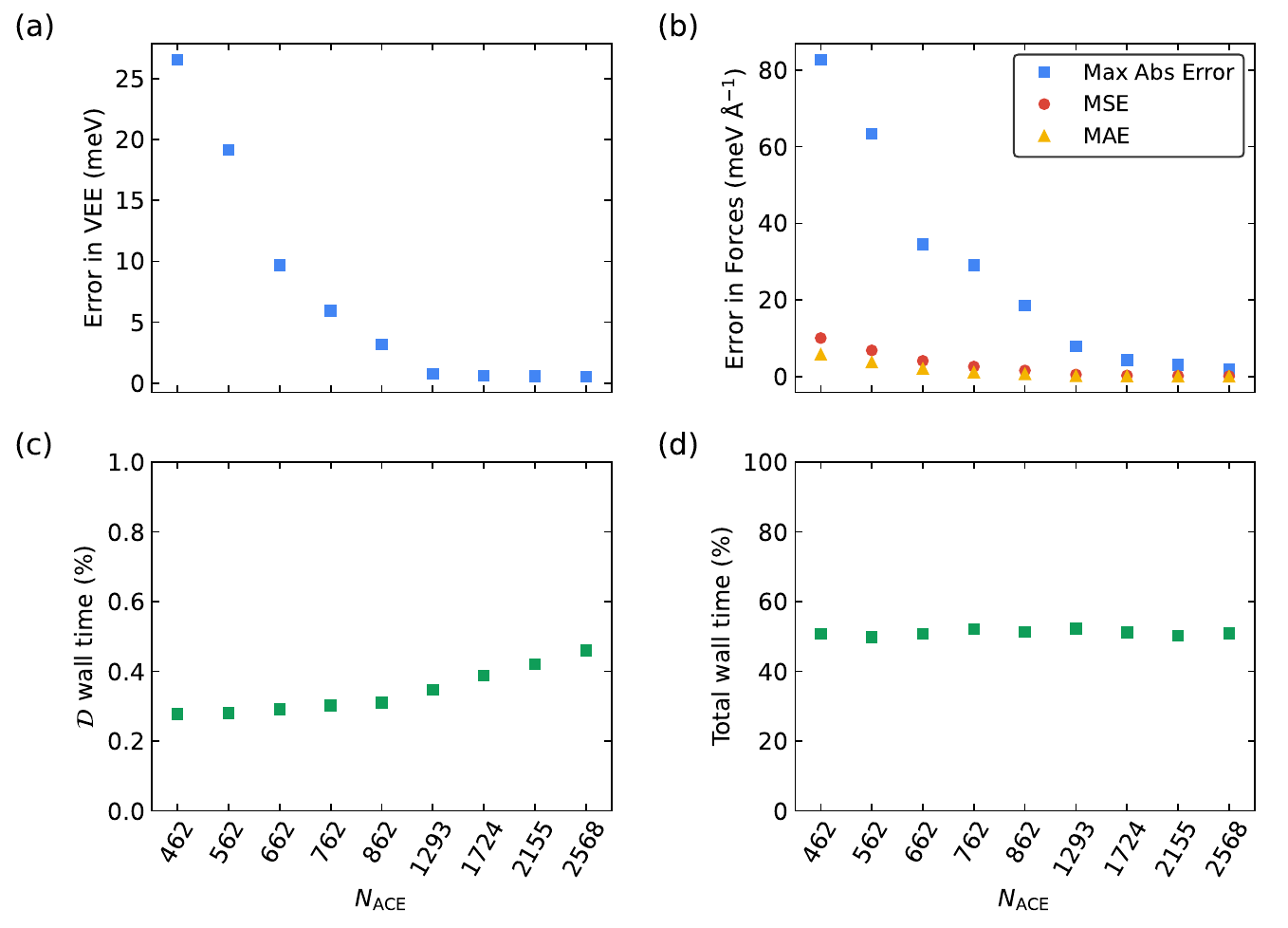}
    \caption{Benchmarks of the use of the adaptively compressed exchange (ACE) operator in TDDFT to speed up the calculation of the $\mathcal{D}$ term of Eq.~\eqref{eq:dterm} as a function of the number of Kohn-Sham orbitals ($N_{\text{ACE}}$) used to build the ACE operator (Eq.~\eqref{eq:ace}). Calculations of the first vertical excitation energy (VEE) and forces of the NV$^-$ in diamond are carried using the DDH hybrid functional, and a conventional $(3 \times 3 \times 3)$ cell with 215 atoms and 432 (430) electrons in the spin-up (spin-down) channel. We report in panels (a) and (b) the error associated with the use of the ACE approximation for the evaluation of the VEE and the forces, respectively. The error is measured considering as a reference the result obtained by directly implementing the exact exchange operator without any approximation. In panels (c) and (d), we report the time savings observed for the calculation of the $\mathcal{D}$ term and the total wall time, respectively. The total wall time is obtained by summing the timings for the calculation of the VEE and the forces. The time savings are reported as the ratio of wall time measured while using the ACE approximation to that measured with no approximation.}
    \label{fig:ACE}
\end{figure}

Figure~\ref{fig:ACE} shows a benchmark of the accuracy and time savings associated with the use of the ACE operator in TDDFT calculations of the VEE and forces for the first excited state of the NV$^-$ in diamond. As expected, errors can be made arbitrarily small by increasing the $N_{\text{ACE}}$ parameter. We also note that we need a smaller $N_{\text{ACE}}$ to converge the calculation of the VEE than that of forces. Setting $N_{\text{ACE},\sigma} = N_{\text{occ},\sigma} + 20$ leads to an error of 27 meV for the VEE, while $N_{\text{ACE},\sigma} = 4 N_{\text{occ},\sigma}$ is required to yield a maximum absolute error (mean absolute error) of 4 meV \AA$^{-1}$ (0.1 meV \AA$^{-1}$) for the forces. The more stringent requirement on forces stems from high-energy unoccupied KS states more likely contributing to the solution of the $Z$-vector equation, Eq.~\eqref{eq:z-v-eq}, than to the calculation of the VEE. The use of the ACE operator allows us to evaluate the $\mathcal{D}$ term at a fraction ($\sim$0.05\%) of the cost of directly applying the exchange operator, yielding a two-fold reduction of the total wall time for the calculation of the VEE and the forces, as shown in Figure~\ref{fig:ACE}(c) and (d). In summary, using the ACE operator to approximate the exact exchange operator can significantly reduce the computational cost of the TDDFT calculations for both VEE and nuclear forces, and, as expected, the error caused by the use of the ACE operator can be arbitrarily reduced by increasing the number of KS orbitals used in its construction.

\subsection{Approximating the $\mathcal{K}^{1d}$ term}
To compute VEEs and analytical forces within TDDFT, the application of the $\mathcal{K}^{1d}$ term (see Eq.~\eqref{eq:k1d}) to an arbitrary set of functions, $\mathcal{A} = \{a_{v\sigma}: v = 1, \cdots, N_{\text{occ},\sigma}; \sigma=\uparrow,\downarrow\}$ is required. This amounts to carrying out $N_{\text{occ},\sigma}^2$ integration operations entering the definition of $\tau_{vv'\sigma}$, i.e., $\tau_{vv'\sigma} (\mathbf{r}) = \int d \mathbf{r'} v_c (\mathbf{r,r'}) \varphi_{v'\sigma}^{\ast} (\mathbf{r}') \varphi_{v\sigma} (\mathbf{r}')$. With the aim of exploiting the near-sightedness principle to reduce the number of integrations, we introduce a unitary transformation of the KS orbitals, i.e., $\varphi_{v\sigma} = \sum_{n=1}^{N_{\text{occ},\sigma}} U_{vn\sigma} \Tilde{\varphi}_{n\sigma}$. Because the $\mathcal{K}^{1d}$ is invariant under a unitary transformation of the KS orbitals, we have
\begin{equation}
\begin{aligned}
    (\mathcal{K}^{1d} \mathcal{A})_{v\sigma} &= \alpha_{\text{EXX}} \int d \mathbf{r'} \mathcal{P}^c_{\sigma} (\mathbf{r,r'}) \sum_{v'=1}^{N_{\text{occ},\sigma}} a_{v'\sigma} (\mathbf{r'}) \tau_{vv'\sigma} (\mathbf{r'}) \\
    &= \alpha_{\text{EXX}} \int d \mathbf{r'} \mathcal{P}^c_{\sigma} (\mathbf{r,r'}) 
    \sum_{m=1}^{N_{\text{occ},\sigma}} U_{vm\sigma} \left\{\sum_{n=1}^{N_{\text{occ},\sigma}} \Tilde{a}_{n\sigma} (\mathbf{r'}) \Tilde{\tau}_{mn\sigma} (\mathbf{r'}) \right\}, \\
\end{aligned}
\label{eq:trunc}
\end{equation}
where we have labeled $\Tilde{a}_{n\sigma} (\mathbf{r'}) = \sum_{v'=1}^{N_{\text{occ},\sigma}} U^{\ast}_{v'n\sigma} a_{v'\sigma} (\mathbf{r'})$ and $\Tilde{\tau}_{mn\sigma} (\mathbf{r'}) = \int \mathrm{d} \mathbf{r''} v_c(\mathbf{r',r''}) \Tilde{\varphi}^{\ast}_{n\sigma}(\mathbf{r''}) \Tilde{\varphi}_{m\sigma}(\mathbf{r''})$ the transformed orbitals and integrals, respectively. Here we consider the unitary transformation from occupied KS orbitals into maximally localized Wannier functions~\cite{gygi2003computation}. 
We define the overlap function, $S_{mn\sigma}$, between Wannier orbitals:
\begin{equation}
    S_{mn\sigma} = \dfrac{\int |\Tilde{\varphi}_{m\sigma}(\mathbf{r})|^2  |\Tilde{\varphi}_{n\sigma} (\mathbf{r})|^2 \mathrm{d} \mathbf{r}}{\sqrt{ \int |\Tilde{\varphi}_{m\sigma} (\mathbf{r})|^4 \mathrm{d} \mathbf{r} \int |\Tilde{\varphi}_{n\sigma} (\mathbf{r'})|^4 \mathrm{d} \mathbf{r'}}}
\end{equation}
We proceed by truncating pairs of non-overlapping localized orbitals in Eq.~\eqref{eq:trunc}, i.e., we approximate $\Tilde{\tau}_{mn\sigma}(\mathbf{r}) = 0$, if the corresponding overlap function is smaller than a preset threshold ($S_{mn\sigma} < S_{\text{thr}}$). In this way, the number of $\Tilde{\tau}_{mn\sigma}(\mathbf{r})$ integrals to be evaluated can be greatly reduced, achieving a linear scaling of the computational workload with respect to the number of orbitals~\cite{nguyen2019finite}.

\begin{figure}
    \centering
    \includegraphics[width=14cm]{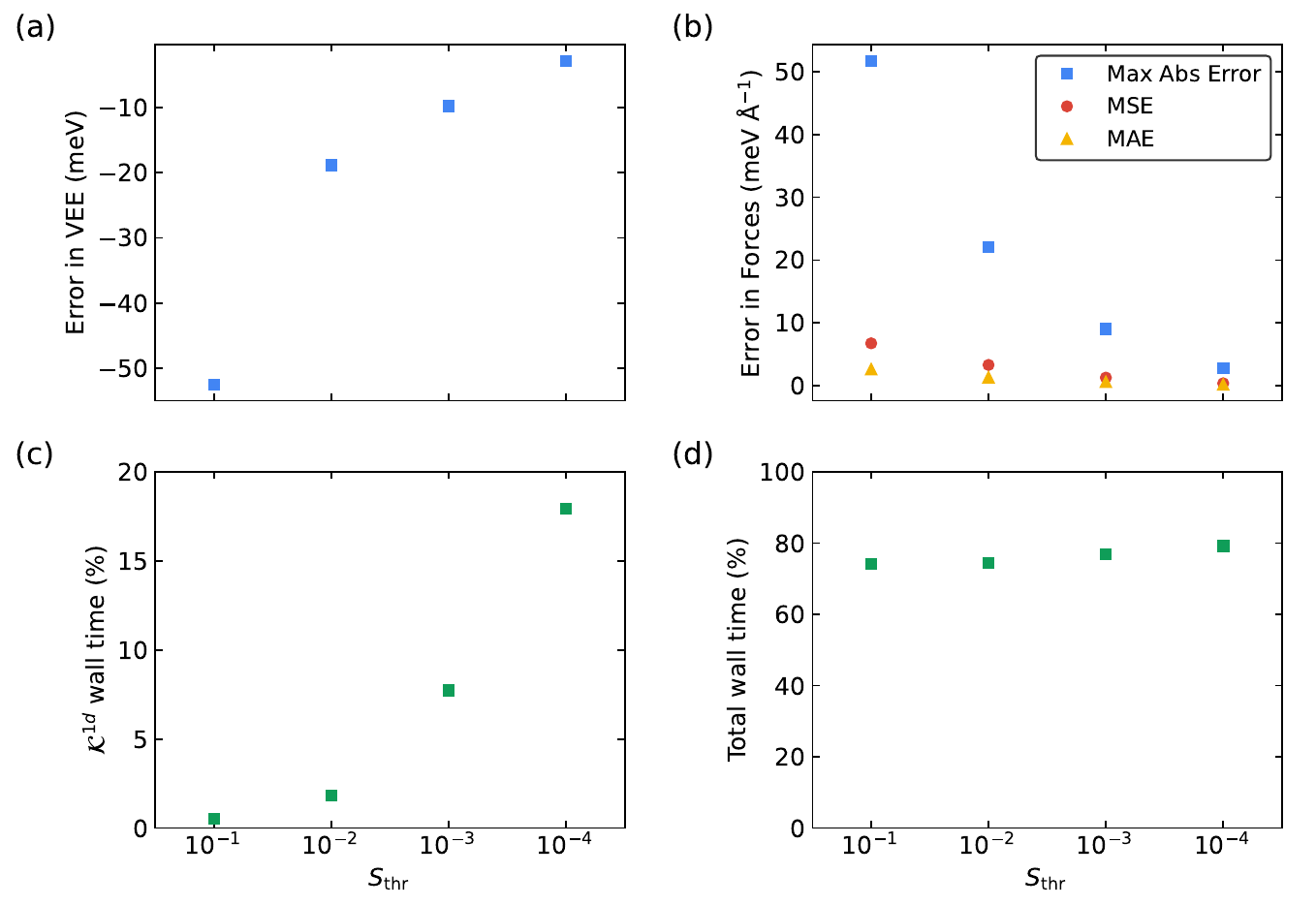}
    \caption{Benchmark of the use of the near-sightedness principle (see text) to approximate the $\mathcal{K}^{1d}$ term of Eq.~\eqref{eq:k1d} in TDDFT calculations, as a function of the overlap threshold $S_{\text{thr}}$. Calculations of the first vertical excitation energy (VEE) and the forces of the NV$^-$ in diamond are carried using the DDH hybrid functional and a conventional $(3 \times 3 \times 3)$ cell with 215 atoms. We report in panels (a) and (b) the error associated with the truncation of pairs of non-overlapping localized orbitals for the evaluation of the VEE and the forces, respectively. In panels (c) and (d), we report the time savings observed for the calculation of the $\mathcal{K}^{1d}$ term and the total wall time, respectively. The total wall time is obtained by summing the timings for the calculation of the energy and the forces. The time savings are reported as the ratio of wall time measured while using truncation to that with no truncation.}
    \label{fig:localization}
\end{figure}

Figure~\ref{fig:localization} shows the accuracy and time savings associated with the use of localized orbitals in TDDFT calculations. Our benchmarks were conducted for the first excited state of the NV$^-$ in diamond. As expected, we can arbitrarily reduce the error by lowering the overlap threshold $S_{\text{thr}}$. For $S_{\text{thr}} = 10^{-3}$, the error in the VEE is only $-10$ meV, and the maximum absolute error and the mean absolute error in nuclear forces is 9 meV \AA$^{-1}$ and 0.6 meV \AA$^{-1}$, respectively. For $S_{\text{thr}} = 10^{-3}$, the number of $\Tilde{\tau}_{mn\sigma}$ integrals, and consequently the computational cost of the $\mathcal{K}^{1d}$ term, is reduced by 92\%, and the total wall time for computing the VEE and the forces is reduced by 23\%, as shown by Figure~\ref{fig:localization}(c) and (d). In summary, by leveraging near-sightedness, we can significantly reduce the computational cost of the TDDFT calculations for the VEE and the nuclear forces, and the error introduced by the approximation can be systematically decreased as a function of the overlap threshold.

\subsection{Approximating the $\mathcal{K}^{2d}$ term}
When computing analytical nuclear forces, the $Z$-vector equation is solved by using the Conjugate-Gradient (CG) algorithm. The implementation requires the repeated application of the super-operator $\mathcal{L} = \mathcal{D} + \mathcal{K}^{1e} - \mathcal{K}^{1d} + \mathcal{K}^{2e} - \mathcal{K}^{2d}$, to residual vectors of Eq.~\eqref{eq:z-v-eq}.  We implemented an inexact Krylov subspace approach~\cite{simoncini2003theory,vandeneshof2004inexact} to accelerate the CG iterations, where the exact $\mathcal{L}$ operator is used until the residual vector becomes smaller than a preset threshold, $\lambda_{\text{thr}}$. Then an approximate operator, $\mathcal{L}_{\text{approx}}$, is applied. We built $\mathcal{L}_{\text{approx}}$ by neglecting the $\mathcal{K}^{2d}$ term in the definition of $\mathcal{L}$, i.e., $\mathcal{L}_{\text{approx}}= \mathcal{D} + \mathcal{K}^{1e} - \mathcal{K}^{1d} + \mathcal{K}^{2e}$.

\begin{figure}
    \centering
    \includegraphics[width=16cm]{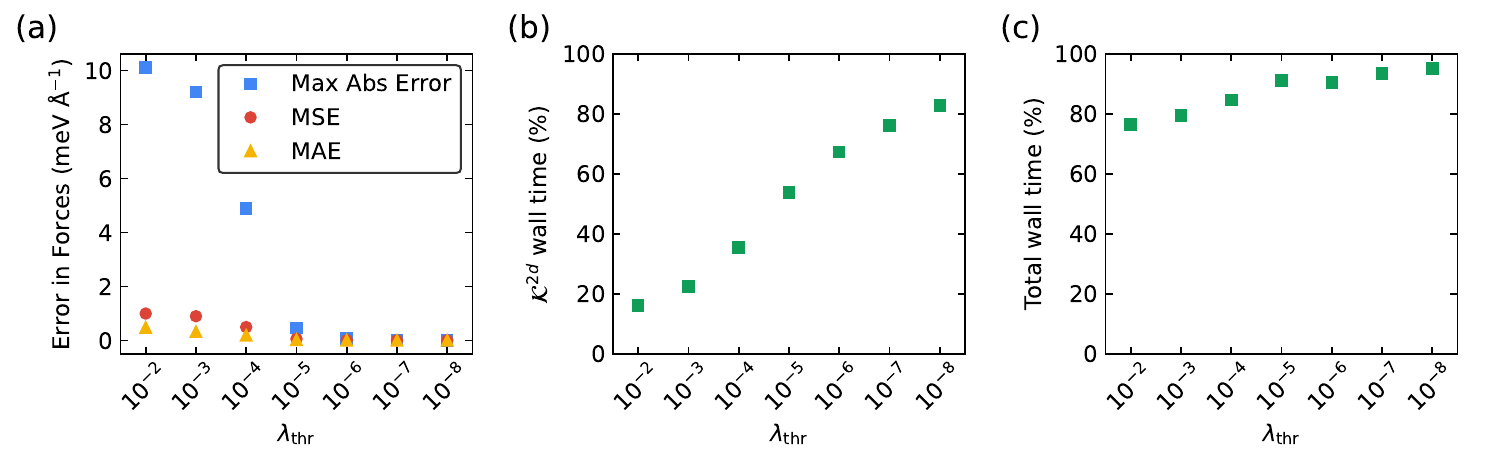}
    \caption{Benchmark of the inexact Krylov subspace approach in TDDFT calculations of analytical nuclear forces with the DDH hybrid functional, as a function of the threshold on the norm of the residual vector $\lambda_{\text{thr}}$ for the calculations of the nuclear forces of the first excited state of the NV$^-$ in diamond. (a) Error in the nuclear forces of the first excited state. (b) The fraction of the wall time for the calculations of the $\mathcal{K}^{2d}$ term of Eq.~\eqref{eq:k2dterm} when using the inexact Krylov subspace approach with respect to using a conventional approach. (c) The fraction of the total wall time of TDDFT calculations using the inexact Krylov subspace approach relative to the total wall time using a conventional approach.}
    \label{fig:inexactKrylov}
\end{figure}

We tested the inexact Krylov subspace approach in TDDFT calculations with the DDH functional for the calculation of the first excited state of the NV$^-$ in diamond, as shown in Figure~\ref{fig:inexactKrylov}. As expected, errors on the nuclear forces decrease as the threshold on the norm of the residual vector $\lambda_{\text{thr}}$ is lowered. With $\lambda_{\text{thr}} = 10^{-4}$, the maximum absolute error and the mean absolute error on nuclear forces is 5 meV \AA$^{-1}$ and 0.2 meV \AA$^{-1}$, respectively, and the number of CG iterations involving operations with the exact $\mathcal{L}$ is reduced from 31 to 11. As a consequence, the computational cost of the $\mathcal{K}^{2d}$ term is reduced by 64\%, and the total wall time for computing VEE and forces is reduced by 15\%, as shown by Figure~\ref{fig:inexactKrylov}(b) and (c). In summary, using the inexact Krylov subspace approach can significantly reduce the computational cost of the TDDFT calculations of nuclear forces, and the error can be systematically reduced by decreasing the threshold on the norm of the residual vector.

\subsection{Using approximations altogether}
\label{subsec:allapproximations}
Finally, we tested the three numerical approximations discussed in the previous subsections altogether in TDDFT calculations with the DDH functional for the calculations of the first excited state of the NV$^-$ in diamond. Compared to the results obtained with the exact TDDFT calculations, when using the three numerical approximations, the error in the VEE is only $-9$ meV, and the maximum absolute error and the mean absolute error in nuclear forces are 10 meV \AA$^{-1}$ and 0.7 meV \AA$^{-1}$, respectively. The total wall time for computing VEE and nuclear forces is reduced by 87\%. The minimal errors indicate that the parameter settings of the three numerical approximations, i.e., $N_{\text{ACE}} = 4 N_{\text{occ}}$ for the ACE operator, $S_{\text{thr}} = 10^{-3}$ for the overlap threshold, and $\lambda_{\text{thr}} = 10^{-4}$ for the threshold on the norm of the residual vector to activate the inexact Krylov subspace approach in the CG solver of the $Z$-vector equation (Eq.~\eqref{eq:z-v-eq}), are reasonable choices. These settings can be an effective starting point when tailoring parameters for other systems. In summary, using the ACE operator, near-sightedness, and the inexact Krylov subspace approach together can significantly reduce the computational cost of the TDDFT calculations of the VEE and the nuclear forces, and the balance between the accuracy and cost can be systematically controlled by adjusting the relevant parameters, as shown above.

\section{Performance and scalability on CPUs and GPUs}
\label{sec:performance}

We implemented spin-conserving and spin-flip TDDFT for the calculation of VEEs and analytical forces in the WEST code~\cite{govoni2015west}, which is a many-body perturbation theory code based on the plane-wave pseudopotential method. In refs.~\citenum{govoni2015west,yu2022west}, we demonstrated the scaling of the code for full-frequency $G_0W_0$ calculations using up to 524,288 CPU cores or 25,920 GPUs. The hierarchical parallelization strategy, which leverages the embarrassingly parallel parts of the algorithms, was proven to be key to achieving excellent scaling. The implementation of WEST-TDDFT adopts a similar multilevel parallelization scheme.

Firstly, in our implementation, the processors are partitioned into subgroups, referred to as images, to facilitate the parallel diagonalization of the super-operator $\mathcal{L}$. The diagonalization is carried out iteratively using the Davidson method~\cite{davidson1975}. Secondly, the processors within each image are partitioned into pools, with each pool computing one spin channel in simulations of spin-polarized systems. Thirdly, the processors within each pool are further partitioned into band groups, with each group being responsible for computing a subset of orbitals. Communications between band groups are required only when a summation over orbitals is carried out, e.g., in Eqs.~\eqref{eq:k1eterm}, \eqref{eq:k2eterm}, \eqref{eq:k1d}, and \eqref{eq:k2dterm}. Finally, as the last parallelization level, the fast Fourier transforms (FFTs) between the direct and reciprocal spaces and the linear algebra operations are carried out using the processors within a band group.

We present an assessment of the performance and scalability of the WEST-TDDFT code on both CPU and GPU nodes using the Perlmutter supercomputer at the National Energy Research Scientific Computing Center (NERSC). Each CPU node of Perlmutter is equipped with two AMD EPYC Milan CPUs. Each GPU node of Perlmutter is equipped with one AMD EPYC Milan CPU and four NVIDIA A100 GPUs. To benchmark the implementation of WEST-TDDFT, we considered the first excited state of the NV$^-$ in diamond, whose ground-state DFT calculation was performed using the Quantum ESPRESSO~\cite{giannozzi2020qe,carnimeo2023quantum} code (version 7.2), the SG15 optimized norm-conserving Vanderbilt (ONCV) pseudopotentials~\cite{hamann2013optimized,schlipf2015optimization}, and the DDH functional~\cite{skone2014ddh,skone2016ddh}. A kinetic energy cutoff of 60 Ry was used for the plane-wave basis set. The Brillouin zone was sampled with the $\Gamma$-point. In the TDDFT calculations, we used the numerical approximations discussed in section~\ref{sec:numericalapproximations}, namely the ACE method with $N_{\text{ACE}} = 4 N_{\text{occ}}$, the Wannier localization with $S_{\text{thr}} = 10^{-3}$, and the inexact Krylov subspace approach with $\lambda_{\text{thr}} = 10^{-4}$.

\begin{figure}
    \centering
    \includegraphics[width=9cm]{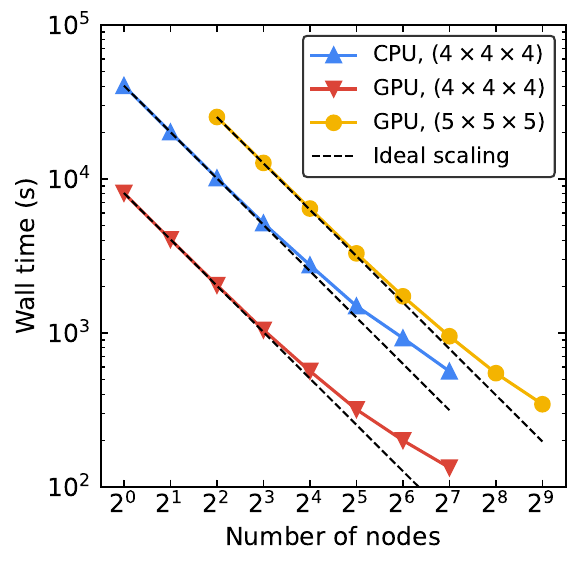}
    \caption{Strong scaling of the WEST-TDDFT code on CPU- and GPU-nodes of the NERSC/Perlmutter supercomputing architecture. The scaling tests report the time to compute the vertical excitation energy (VEE) and the nuclear forces for the first excited state of the NV$^-$ in diamond, simulated using the DDH hybrid functional and 60 Ry kinetic energy cutoff. The blue and red triangles represent timings of simulations in a conventional $(4 \times 4 \times 4)$ supercell of diamond containing 511 atoms, carried out on CPU and GPU nodes, respectively. The orange circles represent the timings of simulations in a conventional $(5 \times 5 \times 5)$ supercell of diamond containing 999 atoms carried out on GPU nodes. The timings presented here amount to the total wall clock time, including the time spent on I/O operations. The black dashed lines indicate the ideal strong scaling.}
    \label{fig:scaling_nv}
\end{figure}

The performance of WEST-TDDFT on CPU- and GPU-nodes is presented in Figure~\ref{fig:scaling_nv} for the NV$^-$ in conventional $(4 \times 4 \times 4)$ and $(5 \times 5 \times 5)$ supercells of diamond containing 511 and 999 atoms, respectively. The figure displays the total wall clock time, including time spent on input/output (I/O) operations. For the $(4 \times 4 \times 4)$ supercell, we observe an excellent strong scaling up to 128 CPU or GPU nodes. The minor degradation of the parallel efficiency is attributed to the overhead caused by the inter-node message-passing interface (MPI) communications and I/O operations that, at a large node count, become comparable to the computational time. For the $(5 \times 5 \times 5)$ supercell, the GPU version of WEST-TDDFT exhibits near-perfect strong scaling up to 512 nodes (2048 GPUs). This showcases the applicability of WEST-TDDFT to conduct large-scale simulations. Further details on the scalability of WEST-TDDFT can be found in section~S3 of the SI.

\newpage
\section{Results}
\label{sec:results}
To validate the implementation of WEST-TDDFT, we computed the VEEs of low-lying excited states of the formaldehyde molecule, and we optimized the geometry of the excited states. We also verified the implementation of the analytical nuclear forces against the calculation of numerical nuclear forces for the excited states of the NV$^-$ in diamond obtained from spin-conserving and spin-flip TDDFT. Details can be found in section~S2 of the SI. Below, we describe results for spin-defects in several solids.

\subsection{Computational setup}
The electronic structures of the defects in diamond,
4H-SiC and MgO are obtained using DFT and the plane-wave pseudopotential method, as implemented in the Quantum ESPRESSO package~\cite{giannozzi2020qe,carnimeo2023quantum}. The plane-wave energy cutoff was set to 60 Ry. We used the semilocal functional by Perdew, Burke, and Ernzerhof (PBE)~\cite{perdew1996generalized} and the DDH functional~\cite{skone2014ddh,skone2016ddh}. The fraction of exact exchange used in the DDH functional was determined by the inverse of the macroscopic dielectric constant of the system, resulting in 18\%, 15\%, and 36\% of exact exchange for diamond, 4H-SiC and MgO, respectively~\cite{skone2014ddh,seo2017designing}.

We used a $(4 \times 4 \times 4)$ supercell containing 511 atoms and a $(5 \times 5 \times 2)$ supercell containing 398 atoms for the triplet excited state of the NV$^-$ in diamond and the VV$^0$ in 4H-SiC, respectively. For singlet excited states, we used a $(3 \times 3 \times 3)$ supercell containing 215 atoms and a $(5 \times 5 \times 1)$ supercell containing 198 atoms for the NV$^-$ in diamond and the VV$^0$ in 4H-SiC, respectively. Supercells of different sizes were used in the study of the finite-size effects for the SiV$^0$ in diamond and V$_{\text{O}}^0$ in MgO. We used the lattice constant optimized with each specific functional~\cite{skone2014ddh,jin2021pl}. The Brillouin zone was sampled with the $\Gamma$ point.

Excited states were computed using the method described in the previous sections. The equilibrium atomic geometries of excited states were obtained by minimizing the nuclear forces below the threshold of 0.01 eV \AA$^{-1}$.

\subsection{NV$^-$ in diamond and VV$^0$ in 4H-SiC}
\begin{figure}
    \centering
    \includegraphics[width=16cm]{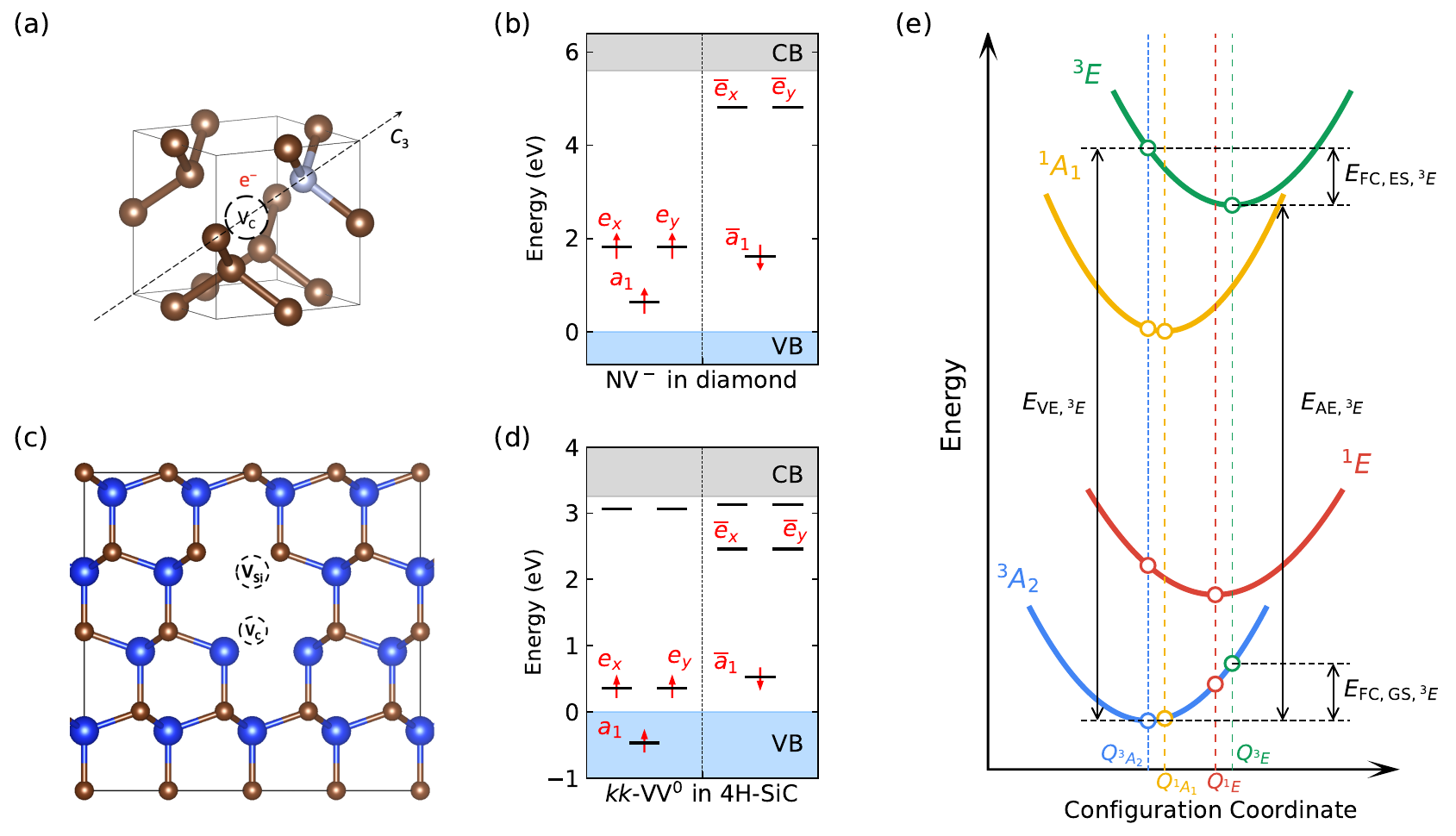}
    \caption{Ball and stick representation of (a) the negatively charged nitrogen-vacancy center (NV$^-$) in diamond and (c) the neutral divacancy center (VV$^0$) in 4H-SiC ($kk$ configuration), with the vacancies displayed as circles in the middle of the cell, and the carbon, nitrogen, and silicon atoms represented by brown, gray and blue spheres, respectively. Position of the single-particle defect levels for (b) NV$^-$ in diamond and (d) VV$^0$ in 4H-SiC, labeled according to the irreducible representation of the $C_{3v}$ group and computed by spin unrestricted density functional theory (DFT) calculations with the DDH hybrid functional. (e) Schematic one-dimensional diagram illustrating potential energy surfaces (PESs) of NV$^-$ in diamond and VV$^0$ in 4H-SiC, including the triplet ground state $^3A_2$, the triplet excited state $^3E$, and singlet states $^1E$ and $^1A_1$. The configuration coordinate of state ${}^3A_2$ is indicated as $Q_{{}^3A_2}$. Vertical excitation energy $(E_{\text{VE}})$ at the ground state geometry, adiabatic excitation energy $(E_{\text{AE}})$, and Franck-Condon shifts in the ground $(E_{\text{FC,GS}})$ and excited state $(E_{\text{FC,ES}})$ PES are indicated by black arrows for the triplet excited state $^3E$. The value of $E_{\text{AE}}$ is frequently used to approximate the energy of the zero-phonon line (ZPL) by assuming that the zero-point energies are similar in the electronic ground and excited state.}
    \label{fig:nvvv}
\end{figure}

The NV$^-$ in diamond and the VV$^0$ in 4H-SiC are prototypical spin defects with numerous quantum technology applications~\cite{davies1976optical,doherty2011negatively,doherty2013nitrogen,gali2019ab,schirhagl2014nitrogen,barry2020sensitivity,christle2017isolated,son2020developing,anderson2022five,li2022room}. In this work we focus on the $kk$ configuration of the VV$^0$ in 4H-SiC, where both the silicon and the carbon vacancy are located in the $k$- site. Although, due to the symmetry of the lattice, there exist three additional configurations of the VV$^0$, the focus here is to describe the applicability and accuracy of the TDDFT techniques developed and implemented in our work, rather than presenting an exhaustive study of spin defects. Both the NV$^-$ in diamond and the VV$^0$ in 4H-SiC have $C_{3v}$ symmetry, and the defect single-particle orbitals within the band gap are $a_1$, and degenerate $e_x$ and $e_y$ orbitals, as shown in Figure~\ref{fig:nvvv}. Many-body electronic states, including the triplet ground state $^3A_2$, the triplet excited state $^3E$, and singlet excited states $^1E$ and $^1A_1$, and the transitions between these states, play an essential role in the operation of the spin defects as qubits, especially in the initialization and readout of the qubit state. Therefore, it is interesting to study their energy and atomic geometries in order to interpret experiments and eventually help design new schemes for quantum technology applications.

We first discuss the many-body electronic excited states of the NV$^-$ in diamond, as obtained using TDDFT. The computed VEEs, adiabatic excitation energies (AEEs), Franck-Condon (FC) shifts, and mass-weighted displacements of the many-body electronic excited states are shown in Table~\ref{tbl:nv}, together with published theoretical and experimental results. Starting from the $M_s = 1$ sublevel of the triplet ground state $^3A_2$ obtained using unrestricted Kohn-Sham DFT, we computed the triplet excited state $^3E$ using spin-conserving TDDFT. The VEEs of state $^3E$ agree within 0.2 eV with those computed using the so-called $\Delta$SCF approach for both PBE and DDH functionals. Unlike $\Delta$SCF, TDDFT calculations do not require any prior knowledge of the symmetry and the composition of the excited state, and they do not suffer from convergence issues often encountered for (near-)degenerate states within the $\Delta$SCF framework. The VEEs of state $^3E$ also agree well, within $\sim$0.2 eV, with the results from the quantum defect embedding theory (QDET)~\cite{sheng2022edc} and with experiments~\cite{davies1976optical}. The small difference between TDDFT and QDET results and experiments may originate, at least in part, from the use of the DDH functional, which uses a screening constant corresponding to that of the pristine crystal. We expect that adopting a hybrid functional that incorporates the screening effects of both the host crystal and the defect~\cite{zhan2023nonempirical} and at the same time taking into account electron--phonon interaction~\cite{yang2021combined,yang2022computational} may improve our TDDFT results. The optimized geometry in the excited state using TDDFT analytical nuclear forces and the resulting FC shifts and displacements are in good agreement with $\Delta$SCF results. The ZPL computed by TDDFT differs from the experimental value by 0.05 (0.17) eV with the PBE (DDH) functional.

\begin{table}
  \caption{Computed vertical excitation energy $E_{\text{VE}}$ (eV), adiabatic excitation energy $E_{\text{AE}}$ (eV), Franck-Condon shift (eV) in the excited ($E_{\text{FC, ES}}$) and the ground states ($E_{\text{FC, GS}}$), and mass-weighted atomic displacement $\Delta Q$ (amu$^{1/2}$ \AA ) for the negatively charged nitrogen-vacancy center NV$^-$ in diamond.}
  \label{tbl:nv}
  \begin{tabular}{lllllll}
    \hline
    Method & $E_{\text{VE}}$ & $E_{\text{AE}}$ & $E_{\text{FC, GS}}$ & $E_{\text{FC, ES}}$ & $\Delta Q$ \\
    \hline
    \addlinespace[5mm]
    & \multicolumn{5}{c}{State $^3E$} \\
    \addlinespace[5mm]
    TDDFT (PBE) & 2.089 & 1.894 & 0.175 & 0.195 & 0.609 \\
    TDDFT (DDH) & 2.372 & 2.112 & 0.228 & 0.256 & 0.659 \\
    $\Delta$SCF (PBE, $\overline{a}_1^1\overline{e}_x^{0.5} \overline{e}_y^{0.5}$ \textsuperscript{\emph{a}} )~\cite{jin2021pl} & 1.937 & 1.731 & 0.180 & 0.206 & 0.620 \\
    $\Delta$SCF (DDH, $\overline{a}_1^1\overline{e}_x^{0.5} \overline{e}_y^{0.5}$)~\cite{jin2021pl} & 2.491 & 2.230 & 0.223 & 0.261 & 0.635 \\
    $\Delta$SCF (PBE, $\overline{a}_1^1\overline{e}_x^{1} \overline{e}_y^{0}$ \textsuperscript{\emph{b}})~\cite{jin2021pl} & & 1.706 &0.203 &  & 0.655 \\
    $\Delta$SCF (DDH, $\overline{a}_1^1\overline{e}_x^{1} \overline{e}_y^{0}$)~\cite{jin2021pl} & & 2.205 & 0.248 & & 0.666 \\
    QDET \textsuperscript{\emph{c}} & 2.162 & & & & \\
    Expt.~\cite{davies1976optical} & 2.18 & 1.945 \textsuperscript{\emph{d}} & & & \\
    \addlinespace[5mm]
    & \multicolumn{5}{c}{State $^1E$} \\
    \addlinespace[5mm]
    TDDFT (PBE) & 0.512 & 0.448 & 0.086 & 0.064 & 0.428 \\
    TDDFT (DDH) & 0.681 & 0.560 & 0.079 & 0.121 & 0.423 \\
    QDET \textsuperscript{\emph{c}} & 0.479 & & & & \\
    Expt.~\cite{rogers2008infrared, kehayias2013infrared, goldman2015state, goldman2015phonon} & & $0.34-0.43$ & & & \\
    \addlinespace[5mm]
    & \multicolumn{5}{c}{State $^1A_1$} \\
    \addlinespace[5mm]
    TDDFT (PBE) & 1.336 & 1.319 & 0.018 & 0.017 & 0.111 \\
    TDDFT (DDH) & 1.973 & 1.957 & 0.018 & 0.016 & 0.095 \\
    QDET \textsuperscript{\emph{c}} & 1.317 & & & & \\
    Expt.~\cite{rogers2008infrared, kehayias2013infrared, goldman2015state, goldman2015phonon} & & $1.51-1.60$ & & & \\
    \hline
  \end{tabular}
  
  \textsuperscript{\emph{a}} The electronic configuration in the hole representation: one hole in the $\overline{a}_1$ orbital, and half a hole in both the $\overline{e}_x$ and the $\overline{e}_y$ orbitals;
  \textsuperscript{\emph{b}} The electronic configuration in the hole representation: one hole in the $\overline{a}_1$ orbital, and one hole in the $\overline{e}_x$ orbital.
  \textsuperscript{\emph{c}} The QDET results are obtained with the exact double counting scheme~\cite{sheng2022edc} and are converged with respect to the active space size.
  \textsuperscript{\emph{d}} The computed adiabatic excitation energy is compared with the experimentally measured zero-phonon line (ZPL) energy~\cite{davies1976optical}.
\end{table}

We also studied the highly correlated singlet excited states $^1E$ and $^1A_1$ using spin-flip TDDFT. The VEEs computed at the DDH level of theory are overestimated compared with those obtained with  QDET and those inferred from experiments. The difference between TDDFT and QDET results may originate from the neglect of double excitations within the TDDFT framework~\cite{jin2022vibrationally}.

In addition to VEE, we show in Table~\ref{tbl:nv} the FC shifts and displacements resulting from the geometry optimizations of the singlet states. The displacements of the $^1A_1$ state are relatively small, as expected since this state has an electronic composition similar to that of the ground state $^3A_2$: one hole in the $a_1$ orbital and two holes in the degenerate $e_x$ and $e_y$ orbitals on average. In contrast, the FC shifts and displacements of the $^1E$ state are larger, although its electronic composition is also not dissimilar from that of the ground state. The reason for that is the coupling with the symmetry broken vibrational modes with $e$ type symmetry. The FC shifts of the $^1E$ state is about half that of the $^3E$ state, indicating a weaker electron--phonon coupling for the ${}^1E \to {}^3A_2$ transition compared to the ${}^3E \to {}^1A_1$ transition, which is consistent with the fact that the experimentally observed multi-phonon inter-system crossing rate is slower for the former process~\cite{robledo2011spin,acosta2010optical,goldman2015phonon}. The optimized geometries of the $^1A_1$ and $^1E$ states have been further validated through the study of vibrationally resolved optical absorption spectra~\cite{jin2022vibrationally}. The computed absorption line shape is in excellent agreement with the experiment~\cite{kehayias2013infrared}, suggesting that spin-flip TDDFT yields an accurate description of the geometries of the highly correlated singlet states.

We now turn to describe the many-body electronic excited states of the VV$^0$ in 4H-SiC; our results are reported in Table~\ref{tbl:vv} together with theoretical and experimental results from previous works. For triplet excited state $^3E$, TDDFT predicts VEEs, AEEs, FC shifts, and displacements in fair agreements with $\Delta$SCF results. Both the TDDFT and the $\Delta$SCF results slightly overestimate the experimental ZPL due to finite-size effects~\cite{davidsson2018first,jin2021pl}. As shown by previous $\Delta$SCF calculations, results obtained with the $(5 \times 5 \times 2)$ supercell used in this work and in the dilute limit differ by $\sim$0.15 eV~\cite{davidsson2018first,jin2021pl}. As for the singlet states $^1E$ and $^1A_1$, spin-flip TDDFT yields an overestimate of their VEEs compared with the constrained random-phase approximation (CRPA) solved by configuration interaction (CI), possibly due to the neglect of double excitations. Unlike the case of the NV$^-$ in diamond, the FC shifts and displacements of the $^1E$ state are comparable with those of the $^3E$ state, indicating a comparable electron-coupling strength and multi-phonon inter-system crossing rate for the ${}^1E \to {}^3A_2$ and the ${}^3E \to {}^1A_1$ transition. These results are consistent with experimental observations for the VV$^0$ in 3C-SiC~\cite{christle2017isolated}, which has a similar electronic structure to that of the VV$^0$ in 4H SiC. In summary, TDDFT can accurately describe the energies and atomic geometries of many-body electronic excited states of both the NV$^-$ in diamond and the VV$^0$ in 4H-SiC and enables the study of multi-phonon processes.

\begin{table}
  \caption{Computed vertical excitation energy $E_{\text{VE}}$ (eV), adiabatic excitation energy $E_{\text{AE}}$ (eV), Franck-Condon shifts (eV) in the excited ($E_{\text{FC, ES}}$) and the ground states ($E_{\text{FC, GS}}$), and mass-weighted atomic displacement $\Delta Q$ (amu$^{1/2}$ \AA ) for the VV$^0$ in 4H-SiC ($kk$ configuration).}
  \label{tbl:vv}
  \begin{tabular}{lllllll}
    \hline
    Method & $E_{\text{VE}}$ & $E_{\text{AE}}$ & $E_{\text{FC, GS}}$ & $E_{\text{FC, ES}}$ & $\Delta Q$ \\
    \hline
    \addlinespace[5mm]
    & \multicolumn{5}{c}{State $^3E$} \\
    \addlinespace[5mm]
    TDDFT (PBE) & 1.413 & & & & \\
    TDDFT (DDH) & 1.464 & 1.351 & 0.097 & 0.113 & 0.779 \\
    $\Delta$SCF (PBE, $\overline{a}_1^1\overline{e}_x^{0.5} \overline{e}_y^{0.5}$ \textsuperscript{\emph{a}} )~\cite{jin2021pl} & 1.233 & 1.161 & 0.063 & 0.072 & 0.635 \\
    $\Delta$SCF (DDH, $\overline{a}_1^1\overline{e}_x^{0.5} \overline{e}_y^{0.5}$)~\cite{jin2021pl} & 1.484 & 1.398 & 0.075 & 0.086 & 0.664 \\
    $\Delta$SCF (PBE, $\overline{a}_1^1\overline{e}_x^{1} \overline{e}_y^{0}$ \textsuperscript{\emph{b}})~\cite{jin2021pl} & & 1.118 & 0.103 & & 0.770 \\
    $\Delta$SCF (DDH, $\overline{a}_1^1\overline{e}_x^{1} \overline{e}_y^{0}$)~\cite{jin2021pl} & & 1.355 & 0.114 & & 0.787 \\
    CI$-$CPRA \textsuperscript{\emph{c}} & 1.13 & & & & & \\
    Expt.~\cite{jin2021pl} & & 1.096 \textsuperscript{\emph{d}} & \\
    \addlinespace[5mm]
    & \multicolumn{5}{c}{State $^1E$} \\
    \addlinespace[5mm]
    TDDFT (PBE) & 0.327 & 0.253 & 0.068 & 0.074 & 0.548 \\
    TDDFT (DDH) & 0.424 & 0.260 & 0.112 & 0.164 & 0.766 \\
    CI$-$CPRA \textsuperscript{\emph{c}} & 0.29 & & & & & \\
    \addlinespace[5mm]
    & \multicolumn{5}{c}{State $^1A_1$} \\
    \addlinespace[5mm]
    TDDFT (PBE) & 0.904 & 0.886 & 0.021 & 0.018 & 0.198 \\
    TDDFT (DDH) & 1.413 & 1.392 & 0.024 & 0.021 & 0.208 \\
    CI$-$CPRA \textsuperscript{\emph{c}} & 0.88 & & & & & \\
    \hline
  \end{tabular}

  \textsuperscript{\emph{a}} The electronic configuration in the hole representation: one hole in the $\overline{a}_1$ orbital, and half a hole in both the $\overline{e}_x$ and the $\overline{e}_y$ orbitals;
  \textsuperscript{\emph{b}} The electronic configuration in the hole representation: one hole in the $\overline{a}_1$ orbital, and one hole in the $\overline{e}_x$ orbital.
  \textsuperscript{\emph{c}} Vertical excitation energies are obtained using the model obtained from constrained random-phase approximation (CRPA) solved by configuration interaction (CI)~\cite{bockstedte2018ab}.
  \textsuperscript{\emph{d}} The computed adiabatic excitation energy is compared with the experimentally measured zero-phonon line (ZPL) energy~\cite{jin2021pl}.
\end{table}

\subsection{SiV$^0$ in diamond}
\begin{figure}
    \centering
    \includegraphics[width=14cm]{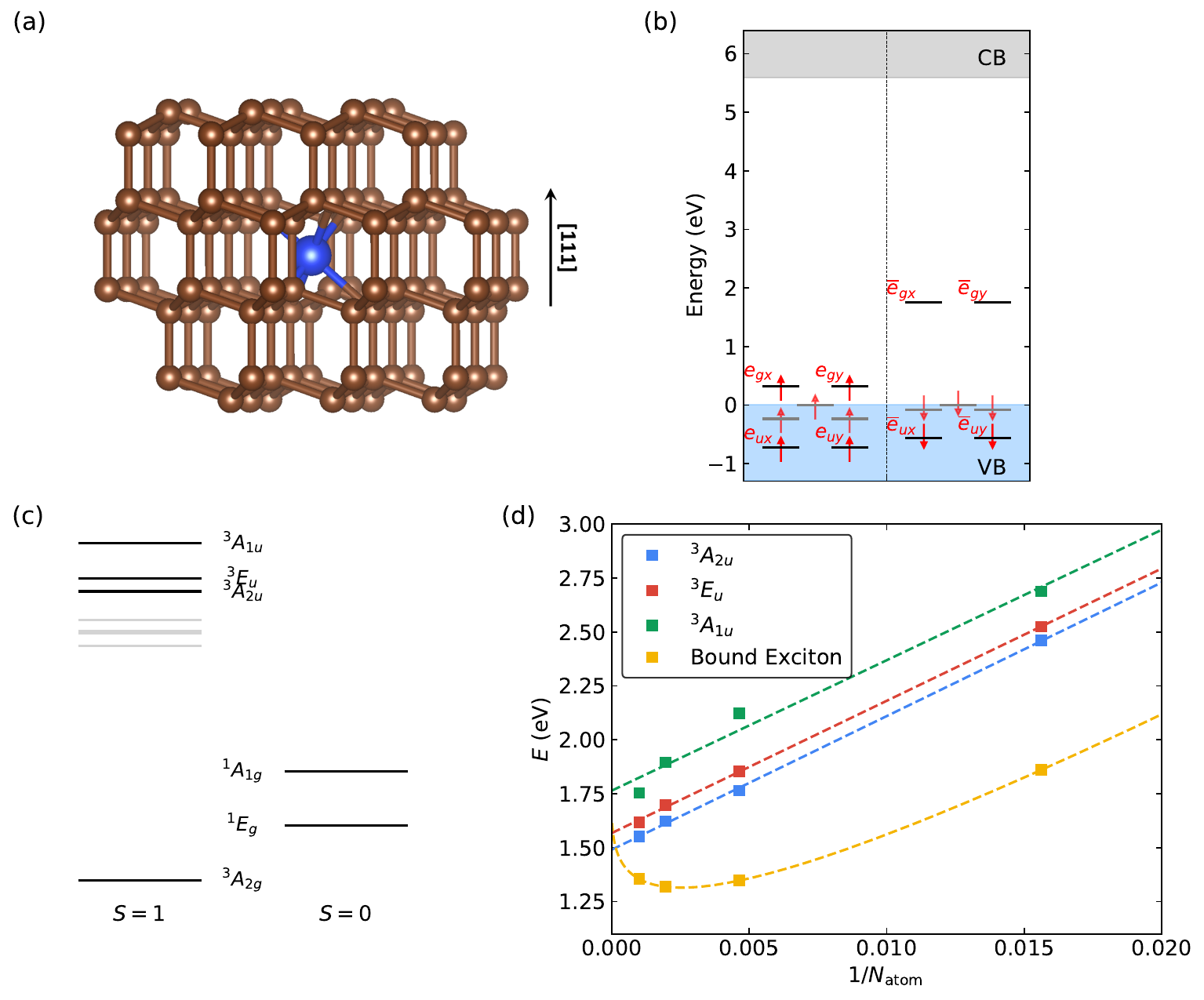}
    \caption{(a) Ball and stick representation of the neutral silicon vacancy center (SiV$^0$) in diamond with the carbon and silicon atoms represented by brown and blue spheres, respectively. (b) Position of the single-particle defect levels for the SiV$^0$ in diamond, labeled according to the irreducible representation of the $D_{3d}$ point group, and computed by spin unrestricted density functional theory (DFT) calculations with the DDH hybrid functional. (c) Ordering of many-body electronic states computed by TDDFT using the DDH functional in a conventional $(4\times4\times4)$ supercell containing 511 atoms. Many-body states resulting from transitions between defect orbitals are shown as solid black lines and labeled according to the irreducible representation of the $D_{3d}$ point group. States resulting from transitions of valence bands (VBs) of diamond into the $\overline{e}_{gx}/\overline{e}_{gy}$ orbital are shown as gray lines and labeled as bound excitons, i.e. an electron in the $\overline{e}_{gx}/\overline{e}_{gy}$ orbital bound to a hole in VBs. (d) Vertical excitation energies of the many-body states $^3A_{2u}$, $^3E_{u}$, and $^3A_{1u}$ from defect excitations and the bound exciton state of the transition from the valence band maximum (VBM) to $\overline{e}_{gx}/\overline{e}_{gy}$ computed using TDDFT and the DDH functional in supercells containing different numbers of atoms. Fitting functions are shown as dashed lines. The VEEs of defect excitations $^3A_{2u}$, $^3E_u$, and $^3A_{1u}$ are fitted by linear functions of $1/N_{\text{atom}}$, while the fitting function Eq.~\eqref{eq:nonlinear} has been used for the VEE of the bound exciton state. Details on the fitting process can be found in section~S5 of the Supporting Information (SI).}
    \label{fig:siv0}
\end{figure}

The SiV$^0$ in diamond exhibits long spin coherence time together with a near-infrared fluorescence signal and has been proposed as a platform for quantum communication applications~\cite{rose2018observation,green2019electronic}. Here, we studied the excited states resulting from the transitions between defect orbitals (so-called \textit{defect excitations}), as well as from transitions between the valence bands (VBs) of diamond and defect orbitals (so-called \textit{bound excitons}). The atomic geometries and electronic structures of the SiV$^0$ are displayed in Figure~\ref{fig:siv0}. This defect possesses $D_{3d}$ symmetry, with the $e_g$ type defect orbitals localized in the band gap and the $e_u$ type defect orbitals resonant with the VBs of diamond. The hybridization of the $e_u$ orbitals with the VBs points to the possible impact of finite-size effects on excited states involving transitions from $e_u$ orbitals into $e_g$ orbitals.

Similar to the NV$^-$ in diamond, SiV$^0$ in diamond has a triplet ground state, $^3A_{2g}$, as shown in Figure~\ref{fig:siv0}, and in the $M_s = 1$ sublevel $e_{gx}$ and $e_{gy}$ orbitals are filled by two electrons, while $\overline{e}_{gx}$ and $\overline{e}_{gy}$ orbitals are empty. Spin conserving excitations from $\overline{e}_{ux}$ and $\overline{e}_{uy}$ to $\overline{e}_{gx}$ and $\overline{e}_{gy}$ yield three triplet excited states $^3A_{2u}$, $^3E_{u}$, and $^3A_{1u}$, which are all linear combinations of the $\overline{e}_{ux} \to \overline{e}_{gx}$, $\overline{e}_{ux} \to \overline{e}_{gy}$, $\overline{e}_{uy} \to \overline{e}_{gx}$, and $\overline{e}_{uy} \to \overline{e}_{gy}$ single excitations~\cite{thiering2019eg}. Spin-conserving TDDFT can correctly describe the multi-configuration nature of the triplet excited states and yield states with the correct symmetry. To estimate finite-size effects on the computed energy of the triplet excited states $^3A_{2u}$, $^3E_{u}$, and $^3A_{1u}$, we performed TDDFT calculations in supercells with different numbers of atoms; our results for the VEEs are summarized in Figure~\ref{fig:siv0}. The linear dependence of the VEEs on $1/N_{\text{atom}}$ stems from the dipole-dipole interaction between the localized excitons in periodic images. In contrast, the VEEs of singlet excited states $^1E_g$ and $^1A_{1g}$ depend weakly on the supercell size since the composition of their respective many-body wavefunctions mainly involve transitions from $e_{gx} (e_{gy})$ orbitals to $\overline{e}_{gx} (\overline{e}_{gy})$ orbitals, which are all within the band gap of diamond (see section~S5 of the SI for details). We summarize the VEEs of both triplet and singlet excited states extrapolated to the dilute limit in Table~\ref{tbl:veesiv0}, together with results from DMET calculations. For triplet states $^3A_{2u}$, $^3E_{u}$, and $^3A_{1u}$ TDDFT predicts much lower VEEs than DMET, closer to the experimental ZPL of 1.31 eV~\cite{d2011optical}. This discrepancy could be attributed to two factors: (i) the neglect of finite-size effects on VEEs in the DMET calculations and (ii) the unsatisfactory treatment of the hybridization between $e_{ux} (e_{uy})$ orbitals and the VBs of diamond in the DMET calculations. The former could reduce the VEE of the triplet excited states by about 0.3 eV, based on the extrapolation of the $(3\times3\times3)$ TDDFT results to the dilute limit. A thorough understanding of reason (ii) requires further explorations, which are beyond the scope of this work.

\begin{table}
  \caption{Computed vertical excitation energies $E_{\text{VE}}$ (eV) for the neutral silicon vacancy center (SiV$^0$) in diamond. TDDFT results reported here are extrapolated to the dilute limit.
  }
  \label{tbl:veesiv0}
  \begin{tabular}{llllllll}
    \hline
    Method & Cell Size & \multicolumn{3}{c}{Triplet} & \multicolumn{2}{c}{Singlet} \\
    & & $^3A_{2u}$ & $^3E_u$ & $^3A_{1u}$ & $^1E_{g}$ & $^1A_{1g}$ \\
    \hline
    TDDFT (PBE) & Dilute Limit & 1.235 & 1.280 & 1.367 & 0.195 & 0.377 \\
    TDDFT (DDH) & Dilute Limit & 1.490 & 1.568 & 1.764 & 0.331 & 0.657 \\
    NEVPT2-DMET(10,12)~\cite{mitra2021excited} & $(3 \times 3 \times 3)$ & 2.39 & 2.47 & 2.61 & 0.51 & 1.14 \\
    \hline
  \end{tabular}
\end{table}

Further, we optimized the geometry of the triplet excited states and approximated the energy of the ZPL as the adiabatic excitation energy by assuming that the zero-point energies are comparable in the electronic ground and excited state. The ZPL obtained in this way is 1.27 eV, in excellent agreement with the experimental value of 1.31 eV~\cite{d2011optical}. The geometry relaxation in the triplet excited states reduces the symmetry from $D_{3d}$ to $C_{2h}$ due to the elongation of two Si$-$C bonds by 0.068 \AA\ and the contraction of the other two by 0.029 \AA. Further details on the geometry relaxation, including the FC shifts, atomic displacements, and Si$-$C bond lengths near the defect center, can be found in section~S4 of the SI. The ability to compute TDDFT analytical nuclear forces also allows for the evaluation of the coupling between the transition dipole moment and the nuclear vibrational motion in the many-body electronic excited states, and thus for the calculation of the vibrationally resolved photoluminescence spectrum, including the contribution of the vibrational modes breaking the inversion symmetry through the Herzberg-Teller effect~\cite{londero2018vibrational}. This investigation will be presented in a forthcoming publication.

While the transitions from the $e_{ux} (e_{uy})$ to the $e_{gx} (e_{gy})$ defect states only involve localized orbitals, the transitions from the VBs of diamond into the $e_{gx} (e_{gy})$ defect states involve more delocalized bound exciton orbitals. Bound excitons have been observed experimentally with excitation energies comparable to the ZPL of the triplet defect excited states~\cite{zhang2020optically}. We computed the VEE of the bound exciton as the transition from the valence band maximum (VBM) to the $\overline{e}_{gx} (\overline{e}_{gy})$ defect states in supercells of different sizes. Our results are summarized in Figure~\ref{fig:siv0}(d). In contrast to the triplet defect excited states, the VEE of the bound exciton state has a non-linear dependence on the supercell size, which can be attributed to the combination of the dipole-dipole interaction of excitons in nearby periodic images and the electron-hole interaction in the same exciton. We fitted the VEE using the function
\begin{equation}
    \label{eq:nonlinear}
    E_{\text{VE}}(L) = E_{\text{VE}}(L = \infty) + \frac{A}{L} \exp \left(-\frac{L}{D}\right) - \frac{B}{L^3},
\end{equation}
where $L \propto N_{\text{atom}}^{1/3}$ is the length of the cubic supercell, $A$ and $B$ are fitting parameters determined by the strength of the electron-hole interaction and the dipole-dipole interaction, respectively. $D$ is a screening parameter, taking into account the fact that the electron-hole interaction no longer depends on $L$, when $L$ is much larger than the radius of the bound exciton. Using $D \in [10, 40]$ \AA\ estimated by a previous study~\cite{zhang2020optically}, we obtain $E_{\text{VE}}(L = \infty) \in [1.33, 1.50]$ eV, higher than the ZPL of the triplet defect excited states, and in fair agreement with the experimental value of 1.39 eV~\cite{zhang2020optically}. Details on the fitting process can be found in section~S5 of the SI. 

In summary, TDDFT with the DDH functional provides a consistent description of localized defect excitations as well as of bound excitons in SiV$^0$ in diamond and predicts their energies in good agreement with experiments. Further, the affordable computational cost of TDDFT calculations enables a thorough investigation of finite-size effects.

\subsection{V$_{\text{O}}^0$ in MgO}
\begin{figure}
    \centering
    \includegraphics[width=14cm]{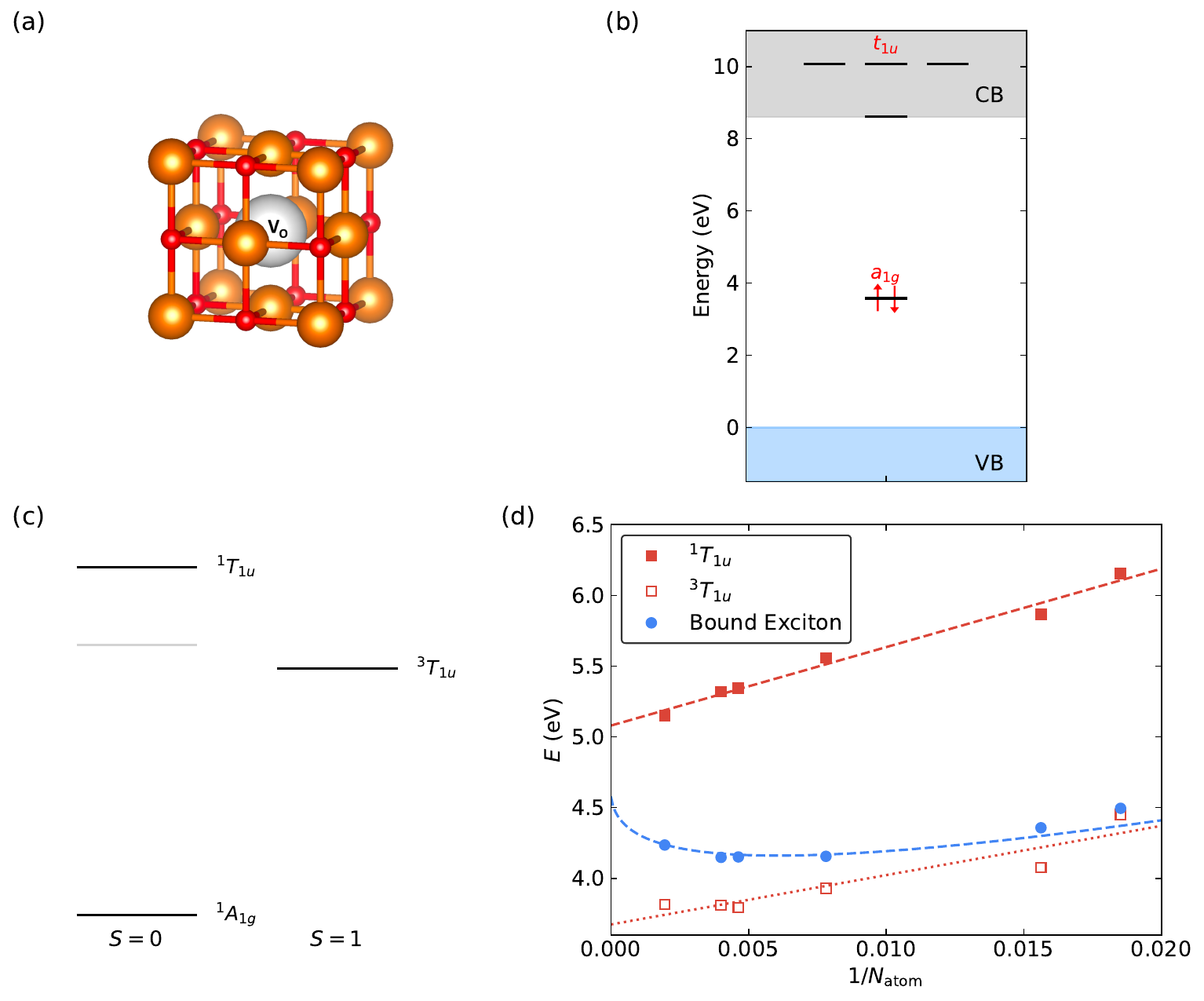}
    \caption{(a) Ball and stick representation of the neutral oxygen vacancy center (V$_{\text{O}}^0$) in magnesium oxide (MgO) with the oxygen and magnesium atoms represented by red and orange spheres, respectively. (b) Position of the single-particle defect levels for V$_{\text{O}}^0$ in MgO, labeled according to the irreducible representation of the $O_{h}$ point group and computed by spin restricted density functional theory (DFT) calculations with the DDH hybrid functional. (c) Ordering of many-body electronic states computed by TDDFT using the DDH functional in a $(3\times3\times3)$ conventional supercell containing 215 atoms. Many-body states resulting from transitions between $a_{1g}$ and $t_{1u}$ defect orbitals are shown as solid black lines and labeled according to the irreducible representation of the $O_{h}$ point group. The state resulting from transitions from the $a_{1g}$ defect orbital into the conduction band minimum (CBM) is shown as a gray line and labeled as bound exciton state, i.e., an electron in the CBM bound to a hole in the $a_{1g}$ orbital. (d) Vertical excitation energies (VEEs) of the many-body excited states computed using TDDFT and the DDH functional in supercells with different numbers of atoms. Fitting functions are shown as dashed lines. The VEEs of defect excitations $^1T_{1u}$ and $^3T_{1u}$ are fitted by linear functions of $1/N_{\text{atom}}$, while the fitting function Eq.~\eqref{eq:nonlinear} is used for the VEE of the bound exciton state. Details on the fitting process can be found in section~S6 of the SI.}
    \label{fig:vo}
\end{figure}

Oxygen vacancies are abundant in MgO and critically influence its applications in spintronic devices~\cite{taudul2017tunneling,schleicher2014localized,miao2008disturbance,velev2007effect} and heterogeneous catalysis~\cite{popov2018multicenter,kulichenko2020periodic,peng1990surface,sharma2011latest,hinuma2018density}. Here we investigate the excited state and optical properties of the neutral oxygen vacancy, V$_{\text{O}}^0$, in MgO using TDDFT, with the aim of interpreting its optical absorption and emission processes. In the ground state atomic geometry, the V$_{\text{O}}^0$ center has $O_h$ symmetry with a $s$-type defect orbital $a_{1g}$ within the band gap and three $p$-type defect orbitals $t_{1u}$ resonant with the conduction bands (CBs) of MgO, as shown in Figure~\ref{fig:vo}. The measured absorption peak at 5.0 eV was attributed to the V$_{\text{O}}^0$ center~\cite{chen1969defect,kappers1970f+}. To interpret the origin of the absorption peak, we computed the VEEs of the low-lying singlet excited states, including the $^1T_{1u}$ state from the $a_{1g} \to t_{1u}$ transition and the bound exciton state from the $a_{1g}$ to the conduction band minimum (CBM) transition. We performed TDDFT calculations with the DDH functional in supercells of different sizes and extrapolated the values of the VEEs to the dilute limit, as shown in Figure~\ref{fig:vo}(d). The VEE of the $^1T_{1u}$ state shows a linear dependence on $1/N_{\text{atom}}$ and reaches 5.08 eV at the dilute limit, which is in good agreement with experiments and with previous theoretical studies using the embedded Bethe--Salpeter equation (BSE)~\cite{vorwerk2023disentangling} and DMET~\cite{verma2023optical} approaches. The ${}^1A_{1g} \to {}^1T_{1u}$ transition is also a dipole-allowed transition with substantial absorption cross section. The bound exciton state, on the other hand, has a VEE depending non-linearly on the system size. We extrapolated the values of the VEE to the dilute limit using Eq.~\eqref{eq:nonlinear}. With $D = \infty$, we obtained a charge transition level of 4.81 eV for the $\text{V}_\text{O}^0 \to \text{V}_\text{O}^+ + e^- ({\text{at CBM}})$ process. With $D \in [21,42]$ \AA\ estimated based on the hydrogenic model, we obtained $E_{\text{VE}}(L = \infty) \in [4.41, 4.58]$ eV, which is smaller than the absorption peak at 5.0 eV. Details on the fitting process can be found in section~S6 of the SI. We note that the bound exciton corresponds to a dipole-forbidden transition, and thus, it is less likely to account for the absorption peak observed experimentally. Therefore, we conclude that the experimentally observed absorption peak can be attributed to the transition from the ground state $^1A_{1g}$ to the localized defect excited state $^1T_{1u}$, in agreement with previous studies.

The emission peaks at $\sim$2.3 to 2.4 eV and $\sim$3.1 to 3.2 eV have been measured experimentally and were suggested to pertain to excitations of the oxygen vacancy in MgO~\cite{chen1969defect,kappers1970f+}. To unravel their origins, we optimized the geometry of different electronic excited states and computed the FC shifts in the ground and the excited state; we then approximated the emission energy as the VEEs at the ground state geometry minus the FC shifts in both the ground and excited states. The computed absorption energies, emission energies, FC shifts, and mass-weighted displacements are summarized in Table~\ref{tbl:relaxvo0}. All excited states exhibit significant FC shifts and displacements as a result of the local excitation around the vacancy site. The emission energy of the $^1T_{1u}$ state is 4.073 eV, much higher than that of the measured emission peaks. Instead, the emission energy of the $^3T_{1u}$ state is 2.824 eV, only 0.4 to 0.5 eV higher than the measured emission energy of $\sim$2.3 to 2.4 eV, consistent with the DMET results~\cite{verma2023optical}. Using a larger supercell for excited-state geometry relaxations would further decrease the emission energy and bring it closer to the experimental value. Therefore, we attribute the experimental emission peak of 2.3 to 2.4 eV to the emission from the $^3T_{1u}$ state. We note that the measured emission peak at 2.3 to 2.4 eV has a relatively long lifetime~\cite{strand2019first}, which is consistent with the slow phosphorescence process from the $^3T_{1u}$ state to the $^1A_{1g}$ ground state. The emission energy of the bound exciton state is higher in energy at $\sim3.04 - 3.21$ eV, and it agrees with the experimental emission peak at 3.1 to 3.2 eV. Therefore, we conclude that the emission peak at 3.1 to 3.2 eV originates from the bound exciton corresponding to the $\text{V}_\text{O}^+ + e^{-} (\text{at CBM}) \to \text{V}_\text{O}^0$ process. Further characterization of the emission peaks would require detailed experimental and theoretical studies on the optical emission spectra.

\begin{table}
   \caption{Computed absorption energy $E_{\text{Abs}}$ (eV), emission energy $E_{\text{Emi}}$, Franck-Condon shifts (eV) in electronic excited states $E_{\text{FC, ES}}$ and electronic ground states $E_{\text{FC, GS}}$, and mass-weighted atomic displacements $\Delta Q$ (amu$^{1/2}$ \AA ) for the neutral oxygen vacancy center (V$_{\text{O}}^0$) in magnesium oxide (MgO). Absorption energies are computed using TDDFT with the DDH hybrid functional and extrapolated to the dilute limit. FC shifts and displacements are computed using TDDFT with the DDH functional and a conventional $(3 \times 3 \times 3)$ supercell with 215 atoms.}
   \label{tbl:relaxvo0}
   \begin{tabular}{lllllll}
     \hline
     States & $E_{\text{Abs}}$ & $E_{\text{Emi}}$ & $E_{\text{FC, GS}}$ & $E_{\text{FC, ES}}$ & $\Delta Q$ \\
     \hline
     ${}^1T_{1u}$ & 5.080 & 4.073 & 0.607 & 0.400 & 1.160 \\
     ${}^3T_{1u}$ & 3.674 & 2.824 & 0.419 & 0.431 & 1.573 \\
     Bound Exciton & $4.41 - 4.58$ & $3.04 - 3.21$ & 0.643 & 0.732 & 1.092 \\
     \hline
   \end{tabular}
\end{table}

In summary, in the case of the oxygen vacancy in MgO TDDFT with the DDH functional provides an accurate treatment of excited state geometry relaxation and enables the study of finite-size effects, thus facilitating the interpretation of optical absorption and emission mechanisms of the V$_\text{O}^0$ in MgO. Our calculations predict substantial geometry displacements and FC shifts for the excited states of the V$_\text{O}^0$ in MgO; such a shift has been frequently neglected in previous theoretical studies.

It is worth noting that unlike point defects in diamond and SiC, where the excited state energies from TDDFT calculations using semilocal and hybrid functionals agree with each other within 0.3 eV, for V$_\text{O}^0$ in MgO, using semilocal functionals can underestimate the excited state energies obtained using hybrid functionals and experiments by more than 1 eV. The FC shifts and displacements can also differ (see Table S5 of the SI for PBE results), depending on the functional. Therefore, it is necessary to use hybrid functionals to study the electronic structure of defects in oxides such as MgO.

\section{Conclusions}
\label{sec:conclusions}
We presented an efficient implementation of spin-conserving and spin-flip TDDFT, including the calculations of analytical nuclear forces based on the plane-wave pseudopotential formalism. Our implementation in the WEST code includes several numerical approximations whose impact has been assessed in detail and a multilevel parallelization strategy that enables a strong scaling up to hundreds of CPU and GPU nodes. The numerical approximations introduced here allow for the study of excited states of systems with thousands of atoms using hybrid functionals. We presented results for the NV$^-$ in diamond and the VV$^0$ in 4H-SiC, showing that spin-conserving and spin-flip TDDFT accurately describe the energy and atomic geometry of triplet and highly-correlated singlet excited states, respectively. In addition, our calculations enable the investigation of multi-phonon processes, including vibrationally resolved optical absorption and emission spectra and inter-system crossing transitions. Our study of the SiV$^0$ in diamond demonstrates that TDDFT is suitable to describe both defect excitations and bound excitons once finite-size effects are properly taken into account by using our highly scalable implementation. Our TDDFT study of the V$_{\text{O}}^0$ in MgO points to the key importance of obtaining accurate excited state geometries to correctly interpret absorption and emission mechanisms. In summary, our implementation of TDDFT with analytical nuclear forces provides a valuable tool to study excited states and optical properties of point defects in a variety of semiconductors and insulators. 

We are exploring the possibility of extending our TDDFT calculations by incorporating additional hybrid functionals, possibly leading to an improvement in the description of screening effects in solids, especially in low-dimensional hosts such as two-dimensional hexagonal boron nitride (2D-hBN)~\cite{zhan2023nonempirical}. Work is in progress to further accelerate TDDFT calculations through the use of density fitting techniques~\cite{hu2020accelerating,qin2023interpolative}, which could enable the direct study of multiple point defects and their interactions in the same supercell. Other interesting future efforts include the implementation of spin-flip TDDFT with the multi-collinear formalism to reduce numerical instabilities~\cite{pu2023noncollinear,li2023noncollinear}, and coupling TDDFT with analytical nuclear forces and molecular dynamics.

Data related to this publication are organized using the Qresp software~\cite{govoni2019qresp} and are available online at \url{https://paperstack.uchicago.edu}.

\begin{acknowledgement}

This work was supported by the computational materials science center Midwest Integrated Center for Computational Materials (MICCoM) . MICCoM is part of the Computational Materials Sciences Program funded by the U.S. Department of Energy, Office of Science, Basic Energy Sciences, Materials Sciences, and Engineering Division through the Argonne National Laboratory, under Contract No. DE-AC02-06CH11357. This research used resources of the National Energy Research
Scientific Computing Center (NERSC), a DOE Office of Science User Facility supported by the Office of Science of the U.S. Department of Energy under Contract No. DE-AC02-05CH11231 using NERSC award ALCC-ERCAP0025950, and resources of the University of Chicago Research Computing Center.

\end{acknowledgement}

\begin{suppinfo}
The Supporting Information is available free of charge at \url{http://pubs.acs.org}.

Additional equations for TDDFT (section~S1), verification and validation of the implementation of WEST-TDDFT (section~S2), details on the scalability of WEST-TDDFT (section~S3), excited-state geometry relaxation of the SiV$^0$ in diamond (section~S4), extrapolation of vertical excitation energies of the SiV$^0$ in diamond (section~S5), and extrapolation of vertical excitation energies of the $\text{V}_{\text{O}}^0$ in MgO (section~S6).
\end{suppinfo}

\bibliography{Main_v1.bib}

\end{document}


\newpage
\tableofcontents
\newpage

\section{Additional Equations for TDDFT}
The expression for $u_{v\sigma}(\mathbf{r})$ for spin-conserving TDDFT within the TDA as defined in Eq.~(20) of the main text reads:
\begin{equation}
\begin{aligned}
    u_{v\sigma} (\mathbf{r}) &= \dfrac{\delta \left[ \mathcal{A}^{\dagger} \left( \mathcal{D} + \mathcal{K}^{1e} - \mathcal{K}^{1d} \right) \mathcal{A}\right]}{\delta \varphi^{\ast}_{v\sigma}(\mathbf{r})} \\
    &= \varphi_{v\sigma} (\mathbf{r}) \sum_{\sigma'} \int \mathrm{d} \mathbf{r'} f_{\text{Hxc},\sigma\sigma'}^{\text{loc}} (\mathbf{r,r'}) \Delta \rho_{\sigma'}^{(x)}(\mathbf{r'}) \\
    &- \alpha_{\text{EXX}} \sum_{v'=1}^{N_{\text{occ},\sigma}} a_{v'\sigma} (\mathbf{r}) \int \mathrm{d} \mathbf{r'} v_c (\mathbf{r,r'}) a^{\ast}_{v'\sigma} (\mathbf{r'}) \varphi_{v\sigma}(\mathbf{r'}) \\
    &+ \alpha_{\text{EXX}} \sum_{v'=1}^{N_{\text{occ},\sigma}} \sum_{v''=1}^{N_{\text{occ},\sigma}} \varphi_{v''\sigma} (\mathbf{r}) \int \mathrm{d} \mathbf{r''}  a^{\ast}_{v'\sigma} (\mathbf{r''}) a_{v''\sigma} (\mathbf{r''}) \int \mathrm{d} \mathbf{r'} v_c(\mathbf{r}, \mathbf{r}') \varphi^{\ast}_{v'\sigma} (\mathbf{r'}) \varphi_{v\sigma} (\mathbf{r'}) \\
    &+ a_{v\sigma} (\mathbf{r}) \sum_{\sigma'} \int \mathrm{d} \mathbf{r'} f_{\text{Hxc},\sigma\sigma'}^{\text{loc}} (\mathbf{r,r'})
    \Delta \rho_{\sigma'}^{\ast} (\mathbf{r'}) \\
    &- \sum_{v'=1}^{N_{\text{occ},\sigma}} a_{v'\sigma}(\mathbf{r}) \int \mathrm{d} \mathbf{r'} \varphi_{v'\sigma}^{\ast} (\mathbf{r'}) \varphi_{v\sigma} (\mathbf{r'}) \sum_{\sigma'} \int \mathrm{d} \mathbf{r''} f_{\text{Hxc},\sigma\sigma'}^{\text{loc}} (\mathbf{r',r''}) \Delta \rho_{\sigma'} (\mathbf{r''}) \\
    &+ \varphi_{v\sigma}(\mathbf{r}) \sum_{\sigma^{\prime}} \sum_{\sigma^{\prime\prime}} \int \mathrm{d} \mathbf{r^{\prime}} \int \mathrm{d} \mathbf{r^{\prime\prime}} g_{\text{xc},\sigma\sigma'\sigma''}(\mathbf{r,r',r''})
    \Delta \rho^{\ast}_{\sigma^{\prime}} (\mathbf{r^{\prime}}) \Delta \rho_{\sigma^{\prime\prime}} (\mathbf{r^{\prime\prime}}) \\
    &- \alpha_{\text{EXX}} \sum_{v'=1}^{N_{\text{occ},\sigma}} \varphi_{v'\sigma} (\mathbf{r}) \int \mathrm{d} \mathbf{r'} v_c (\mathbf{r,r'}) a_{v'\sigma}^{\ast}(\mathbf{r'}) a_{v\sigma} (\mathbf{r'}) \\
    &+ \alpha_{\text{EXX}} \sum_{v'=1}^{N_{\text{occ},\sigma}} a_{v'\sigma} (\mathbf{r}) \int \mathrm{d} \mathbf{r'} \sum_{v''=1}^{N_{\text{occ},\sigma}} a^{\ast}_{v''\sigma} (\mathbf{r'}) \varphi_{v\sigma} (\mathbf{r'}) \int \mathrm{d} \mathbf{r''} v_c (\mathbf{r',r''}) \varphi^{\ast}_{v'\sigma} (\mathbf{r''}) \varphi_{v''\sigma} (\mathbf{r''}). \\
\end{aligned}
\label{s-eq:Zvectoreq}
\end{equation}

In Eq.~\eqref{s-eq:Zvectoreq}, $\Delta \rho_{\sigma} (\mathbf{r}) = \sum_{v = 1}^{N_{\text{occ},\sigma}} \varphi_{v\sigma}^{\ast} (\mathbf{r}) a_{v\sigma} (\mathbf{r})$, and
\begin{equation}
    g_{\text{xc},\sigma\sigma'\sigma''}^{\text{loc}} (\mathbf{r,r',r''}) = \left.\dfrac{\delta^2 V_{\text{xc},\sigma}^{\text{loc}}(\mathbf{r})}{\delta \rho_{\sigma'}(\mathbf{r'}) \delta \rho_{\sigma''} (\mathbf{r''})} \right |_{\left(\rho^0, \nabla\rho^0 \right)}
\end{equation}
is the second order functional derivative of the exchange-correlation potential with respect to the electron density. The term involving $g_{\text{xc},\sigma\sigma'\sigma''}^{\text{loc}} (\mathbf{r,r',r''})$ can be evaluated by using a finite difference approach and written in the following way:
\begin{equation}
\begin{aligned}
    &\sum_{\sigma^{\prime}} \sum_{\sigma^{\prime\prime}} \int \mathrm{d} \mathbf{r^{\prime}} \int \mathrm{d} \mathbf{r^{\prime\prime}} g^{\text{loc}}_{\text{xc},\sigma\sigma'\sigma''}(\mathbf{r,r',r''})
    \Delta \rho^{\ast}_{\sigma^{\prime}} (\mathbf{r^{\prime}}) \Delta \rho_{\sigma^{\prime\prime}} (\mathbf{r^{\prime\prime}}) \\
    \approx &\sum_k \dfrac{c_k}{12\epsilon^2} V^{\text{loc}}_{\text{xc},\sigma} \left[ \rho^0_{\uparrow} + k\epsilon \Delta \rho_{\uparrow}, \rho^0_{\downarrow} + k\epsilon \Delta \rho_{\downarrow} \right] (\mathbf{r}),
\end{aligned}
\end{equation}
where $c_k = -1, 16, -30, 16, -1$ for $k = -2, -1, 0, 1, 2$. Numerical tests have shown that stable results can be achieved by using $\epsilon = 0.01$~\cite{zhang2015subspace}.

The LR-TDDFT eigenvalue equation in the presence of spin-flip (SF) excitations reads
\begin{equation}
    \left(
    \begin{matrix}
    \mathcal{D}^{\text{SF}} + \mathcal{K}^{1e,\text{SF}} - \mathcal{K}^{1d,\text{SF}} & \mathcal{K}^{2e,\text{SF}} - \mathcal{K}^{2d,\text{SF}} \\
    \mathcal{K}^{2e,\text{SF}} - \mathcal{K}^{2d,\text{SF}} & \mathcal{D}^\text{SF} + \mathcal{K}^{1e,\text{SF}} - \mathcal{K}^{1d,\text{SF}} \\
    \end{matrix}
    \right) \left(
    \begin{matrix}
    \mathcal{A}_s^{\text{SF}} \\
    \mathcal{B}_s^{\text{SF}} \\
    \end{matrix}
    \right) = \omega_s \left(
    \begin{matrix}
    \mathcal{I} & 0 \\
    0 & -\mathcal{I}
    \end{matrix}
    \right) \left(
    \begin{matrix}
    \mathcal{A}_s^{\text{SF}} \\
    \mathcal{B}_s^{\text{SF}} \\
    \end{matrix}
    \right),
\label{eq:TDDFT_VEXX_TDA_SF}
\end{equation}
where $\mathcal{A}_s^{\text{SF}} = \left\{| a^{\text{SF}}_{s,v\sigma}\rangle: v=1,\dots, N_{\text{occ},\widetilde{\sigma}}; \sigma = \uparrow, \downarrow \right\}$ and $\mathcal{B}_s^{\text{SF}} = \left\{| b^{\text{SF}}_{s,v\sigma}\rangle: v=1,\dots, N_{\text{occ},\widetilde{\sigma}}; \sigma = \uparrow, \downarrow \right\}$ are two sets of orbitals that enter the definition of the linear change of the density matrix with respect to the ground state density matrix, due to the $s$-th SF excitation:
\begin{equation*}
    \Delta \rho_{s,\sigma}^{\text{SF}} = \sum_{v=1}^{N_{\text{occ},\widetilde{\sigma}}}  |a^{\text{SF}}_{s,v\sigma} \rangle \langle \varphi_{v\widetilde{\sigma}} | + \sum_{v=1}^{N_{\text{occ},\sigma}} | \varphi_{v\sigma}\rangle \langle b^{\text{SF}}_{s,v\widetilde{\sigma}} |.
\end{equation*}
The operators on the left-hand side (LHS) of Eq.~\eqref{eq:TDDFT_VEXX_TDA_SF} are defined as
\begin{equation}
    \mathcal{D}^{\text{SF}} \mathcal{A}_s^{\text{SF}} = \left\{ \mathcal{P}_{\sigma}^c \left(H_{\sigma}^{\text{KS}} - \varepsilon_{v\widetilde{\sigma}} \right) | a^{\text{SF}}_{s,v\sigma}\rangle : v = 1, \dots, N_{\text{occ},\widetilde{\sigma}}; \sigma = \uparrow, \downarrow \right\},
\end{equation}
\begin{equation}
\begin{aligned}
    &\mathcal{K}^{1e,\text{SF}}\mathcal{A}_s^{\text{SF}} = \\
    &\left\{ \int \mathrm{d} \mathbf{r}^{\prime} \mathcal{P}^c_{\sigma}(\mathbf{r,r'}) \varphi_{v\widetilde{\sigma}}(\mathbf{r'}) \int \mathrm{d} \mathbf{r''} f_{\text{xc}}^{\text{loc,SF}}(\mathbf{r',r''}) \sum_{v'=1}^{N_{\text{occ},\widetilde{\sigma}}} \varphi_{v'\widetilde{\sigma}}^{\ast} (\mathbf{r}'') a^{\text{SF}}_{s,v'\sigma}(\mathbf{r}'') : v = 1, \dots, N_{\text{occ},\widetilde{\sigma}}; \sigma = \uparrow, \downarrow \right\},\\
\end{aligned}
\end{equation}
\begin{equation}
\begin{aligned}
    &\mathcal{K}^{2e,\text{SF}}\mathcal{A}_s^{\text{SF}} = \\
    &\left\{ \int \mathrm{d} \mathbf{r}^{\prime} \mathcal{P}^c_{\sigma}(\mathbf{r,r'}) \varphi_{v\widetilde{\sigma}}(\mathbf{r'}) \int \mathrm{d} \mathbf{r''} f_{\text{xc}}^{\text{loc,SF}}(\mathbf{r',r''}) \sum_{v'=1}^{N_{\text{occ},\sigma}} a_{s,v'\widetilde{\sigma}}^{\text{SF}\ast} (\mathbf{r}'') \varphi_{v'\sigma}(\mathbf{r}'') : v = 1, \dots, N_{\text{occ},\widetilde{\sigma}}; \sigma = \uparrow, \downarrow \right\},\\
\end{aligned}
\end{equation}
\begin{equation}
\begin{aligned}
    &\mathcal{K}^{1d,\text{SF}}\mathcal{A}^{\text{SF}}_s = \\
    &\bigg\{ \alpha_{\text{EXX}} \int \mathrm{d} \mathbf{r'} \mathcal{P}^c_{\sigma}(\mathbf{r,r'}) &\sum_{v'=1}^{N_{\text{occ},\widetilde{\sigma}}} a^{\text{SF}}_{s,v'\sigma} (\mathbf{r'}) \int \mathrm{d} \mathbf{r''} v_c(\mathbf{r',r''}) \varphi^{\ast}_{v'\widetilde{\sigma}}(\mathbf{r''}) \varphi_{v\widetilde{\sigma}}(\mathbf{r''}): v = 1, \dots, N_{\text{occ},\widetilde{\sigma}}; \sigma = \uparrow, \downarrow \bigg\}.\\
\end{aligned}
\end{equation}
\begin{equation}
\begin{aligned}
    &\mathcal{K}^{2d,\text{SF}}\mathcal{A}_s^{\text{SF}} = \\
    &\bigg\{ \alpha_{\text{EXX}} \int \mathrm{d} \mathbf{r'} \mathcal{P}^c_{\sigma}(\mathbf{r,r'}) &\sum_{v'=1}^{N_{\text{occ},\sigma}} \varphi_{s,v'\sigma} (\mathbf{r'}) \int \mathrm{d} \mathbf{r''} v_c(\mathbf{r',r''}) a^{\text{SF}\ast}_{v'\widetilde{\sigma}}(\mathbf{r''}) \varphi_{v\widetilde{\sigma}}(\mathbf{r''}): v = 1, \dots, N_{\text{occ},\widetilde{\sigma}}; \sigma = \uparrow, \downarrow \bigg\}.\\
\end{aligned}
\end{equation}
The SF exchange-correlation kernel is defined in Eq.~(26) of the main text. Under the TDA, the $\mathcal{K}^{2e,\text{SF}}$ and $\mathcal{K}^{2d,\text{SF}}$ terms in Eq.~\eqref{eq:TDDFT_VEXX_TDA_SF} are neglected, yielding $\mathcal{B}_s^{\text{SF}} = 0$, and one solves the eigenvalue problem as defined in Eq.~(21) of the main text.

The expression for $u_{v\sigma}(\mathbf{r})$ for spin-flip TDDFT within the TDA reads:
\begin{equation}
\begin{aligned}
    u_{v\sigma} (\mathbf{r}) &= \dfrac{\delta \left[ \mathcal{A}^{\text{SF}\dagger} \left( \mathcal{D}^{\text{SF}} + \mathcal{K}^{1e,\text{SF}} - \mathcal{K}^{1d,\text{SF}} \right) \mathcal{A}^\text{SF} \right]}{\delta \varphi^{\ast}_{v\sigma}(\mathbf{r})} \\
    &= \varphi_{v\sigma} (\mathbf{r}) \sum_{\sigma'} \int \mathrm{d} \mathbf{r'} f_{\text{Hxc},\sigma\sigma'}^{\text{loc}} (\mathbf{r,r'}) \Delta \rho_{\sigma'}^{(x)}(\mathbf{r'}) \\
    &- \alpha_{\text{EXX}} \sum_{v'=1}^{N_{\text{occ},\widetilde{\sigma}}} a^{\text{SF}}_{v'\sigma} (\mathbf{r'}) \int \mathrm{d} \mathbf{r'} v_c (\mathbf{r,r'}) a^{\text{SF}\ast}_{v'\sigma} (\mathbf{r'}) \varphi_{v\sigma}(\mathbf{r'}) \\
    &+ \alpha_{\text{EXX}} \sum_{v'=1}^{N_{\text{occ},\sigma}} \sum_{v''=1}^{N_{\text{occ},\sigma}} \varphi_{v''\sigma} (\mathbf{r}) \int \mathrm{d} \mathbf{r''}  a^{\text{SF}\ast}_{v'\widetilde{\sigma}} (\mathbf{r''}) a^{\text{SF}}_{v''\widetilde{\sigma}} (\mathbf{r''}) \int \mathrm{d} \mathbf{r'} v_c(\mathbf{r}, \mathbf{r}') \varphi^{\ast}_{v'\sigma} (\mathbf{r'}) \varphi_{v\sigma} (\mathbf{r'}) \\
    &+ a^{\text{SF}}_{v\widetilde{\sigma}} (\mathbf{r}) \int \mathrm{d} \mathbf{r'} f_{\text{xc}}^{\text{loc,SF}} (\mathbf{r,r'})
    \Delta \rho_{\widetilde{\sigma}}^{\text{SF} \ast} (\mathbf{r'}) \\
    &- \sum_{v'=1}^{N_{\text{occ},\widetilde{\sigma}}} a^{\text{SF}}_{v'\sigma}(\mathbf{r}) \int \mathrm{d} \mathbf{r'} \varphi_{v'\widetilde{\sigma}}^{\ast} (\mathbf{r'}) \varphi_{v\sigma} (\mathbf{r'}) \int \mathrm{d} \mathbf{r''} f_{\text{xc}}^{\text{loc,SF}} (\mathbf{r',r''}) \Delta \rho_{\sigma}^{\text{SF}} (\mathbf{r''}) \\
    &+ \varphi_{v\sigma}(\mathbf{r}) \int \mathrm{d} \mathbf{r^{\prime}} \int \mathrm{d} \mathbf{r^{\prime\prime}} g_{\text{xc},\sigma}^{\text{loc,SF}}(\mathbf{r,r',r''}) \sum_{\sigma'}
    \Delta \rho^{\text{SF}\ast}_{\sigma'} (\mathbf{r^{\prime}}) \Delta \rho_{\sigma'}^{\text{SF}} (\mathbf{r^{\prime\prime}}) \\
    &- \alpha_{\text{EXX}} \sum_{v'=1}^{N_{\text{occ},\sigma}} \varphi_{v'\sigma} (\mathbf{r}) \int \mathrm{d} \mathbf{r'} v_c (\mathbf{r,r'}) a_{v'\widetilde{\sigma}}^{\text{SF}\ast}(\mathbf{r'}) a^{\text{SF}}_{v\widetilde{\sigma}} (\mathbf{r'}) \\
    &+ \alpha_{\text{EXX}} \sum_{v'=1}^{N_{\text{occ},\widetilde{\sigma}}} a^{\text{SF}}_{v'\sigma} (\mathbf{r}) \int \mathrm{d} \mathbf{r'} \sum_{v''=1}^{N_{\text{occ},\widetilde{\sigma}}} a^{\text{SF}\ast}_{v''\sigma} (\mathbf{r'}) \varphi_{v\sigma} (\mathbf{r'}) \int \mathrm{d} \mathbf{r''} v_c (\mathbf{r',r''}) \varphi^{\ast}_{v'\widetilde{\sigma}} (\mathbf{r''}) \varphi_{v''\widetilde{\sigma}} (\mathbf{r''}). \\
\end{aligned}
\end{equation}
Here $\Delta \rho_{\sigma}^{\text{SF}} (\mathbf{r}) = \sum_{v=1}^{N_{\text{occ},\widetilde{\sigma}}} \varphi^{\ast}_{v\widetilde{\sigma}} (\mathbf{r}) a^{\text{SF}}_{v\sigma} (\mathbf{r})$. We have used the following definition for the functional derivative of the spin-flip exchange-correlation kernel with respect to the
electron density~\cite{seth2011time,bernard2012general}
\begin{equation}
\begin{aligned}
    &g_{\text{xc},\sigma}^{\text{loc,SF}} (\mathbf{r,r^{\prime},r^{\prime\prime}}) = \left\{
    \begin{aligned}
        &\left.- \dfrac{ V_{\text{xc},\uparrow}^{\text{loc}}(\mathbf{r''}) - V_{\text{xc},\downarrow}^{\text{loc}}(\mathbf{r''}) }{\left[ \rho_{\uparrow}(\mathbf{r''}) - \rho_{\downarrow}(\mathbf{r''}) \right]^2} \right |_{\left(\rho^0, \nabla\rho^0\right)} \delta (\mathbf{r'', r}) \delta \left(\mathbf{r'', r'} \right) \\
        &\qquad + \left.\dfrac{ f_{\text{xc},\uparrow\uparrow}^{\text{loc}}(\mathbf{r'',r}) - f_{\text{xc},\downarrow\uparrow}^{\text{loc}}(\mathbf{r'',r}) }{\rho_{\uparrow}(\mathbf{r''}) - \rho_{\downarrow}(\mathbf{r''})} \right|_{\left( \rho^0, \nabla\rho^0 \right)} \delta \left(\mathbf{r'', r^{\prime}} \right), \quad \sigma = \uparrow \\
        &\left.\dfrac{ V_{\text{xc},\uparrow}^{\text{loc}}(\mathbf{r''}) - V_{\text{xc},\downarrow}^{\text{loc}}(\mathbf{r''}) }{\left[ \rho_{\uparrow}(\mathbf{r''}) - \rho_{\downarrow}(\mathbf{r''}) \right]^2} \right|_{\left(\rho^0, \nabla\rho^0 \right)} \delta (\mathbf{r'', r}) \delta \left(\mathbf{r'', r'} \right) \\ &\qquad + \left.\dfrac{ f_{\text{xc},\uparrow\downarrow}^{\text{loc}}(\mathbf{r'',r}) - f_{\text{xc},\downarrow\downarrow}^{\text{loc}}(\mathbf{r'',r}) }{\rho_{\uparrow}(\mathbf{r''}) - \rho_{\downarrow}(\mathbf{r''})}\right|_{\left(\rho^0, \nabla\rho^0\right)} \delta \left(\mathbf{r'', r^{\prime}} \right), \quad \sigma = \downarrow \\
    \end{aligned}
    \right..
\end{aligned}
\end{equation}

\section{Verification and Validation of the Implementation of WEST-TDDFT}
\subsection{Formaldehyde Molecule}
We verified the implementation of time-dependent density functional theory (TDDFT) in the WEST code (referred to as WEST-TDDFT hereafter) by applying it to the formaldehyde molecule, which is routinely used as a benchmark for TDDFT. Geometry relaxation was performed in excited states ${}^1A^{\prime\prime}$, ${}^1B_2$, and ${}^3A^{\prime\prime}$ with the Tamm-Dancoff approximation (TDA), and the resulting atomic geometries and the adiabatic excitation energies are summarized in Table~\ref{s-tbl:Formaldehyde} and compared with the TDDFT-TDA results reported in literature~\cite{hutter2003excited}. The Quantum ESPRESSO code~\cite{giannozzi2020qe,carnimeo2023quantum} was employed for the ground-state DFT calculations, with the PBE functional~\cite{perdew1996generalized} and the SG15 optimized norm-conserving Vanderbilt (ONCV) pseudopotentials~\cite{hamann2013optimized,schlipf2015optimization}. The dimensions of the supercell were set to 20 \AA $\times$ 20 \AA $\times$ 20 \AA. The energy cutoff of the plane-wave basis was set to 90 Ry. The force convergence criterion for geometry relaxation was set as 0.01 eV \AA$^{-1}$. As shown in Table~\ref{s-tbl:Formaldehyde}, the results obtained using WEST-TDDFT agree very well with the TDDFT results reported in literature~\cite{hutter2003excited}.
\begin{table}
   \caption{Equilibrium structures and adiabatic excitation energies, denoted as $E_{\text{AE}}$ (in eV), for formaldehyde were calculated using WEST-TDDFT and compared with values from the literature~\cite{hutter2003excited}. Both sets of calculations employed the Tamm-Dancoff approximation (TDA) and the PBE functional. Bond lengths are provided in \AA, while angles are in degrees. $\Phi$ represents the out-of-plane angle.}
   \label{s-tbl:Formaldehyde}
   \begin{tabular}{lllllll}
     \hline
     \addlinespace[1mm]
     State & $R_{\text{CO}}$ & $R_{\text{CH}}$ & $\angle_{\text{HCH}}$ & $\Phi$ & $E_{\text{AE}}$ & Reference \\
     \addlinespace[1mm]
     \hline
     \addlinespace[4mm]
     \multirow{2}{*}{${}^1A_1$ (Ground state)}& 1.2065 & 1.1162 & 115.98 & 0 & & WEST-TDDFT \\
     & 1.211 & 1.118 & 116.1 & 0 & & Ref.~\citenum{hutter2003excited} \\
     \addlinespace[4mm]
     \multirow{2}{*}{${}^1A^{\prime\prime}$}& 1.3040 & 1.1009 & 117.13 & 30.63 & 3.544 & WEST-TDDFT \\
     & 1.308 & 1.103 & 116.8 & 30.0 & 3.53 & Ref.~\citenum{hutter2003excited} \\
     \addlinespace[4mm]
     \multirow{2}{*}{${}^1B_2$}& 1.2019 & 1.1133 & 122.03 & 0 & 5.776 & WEST-TDDFT \\
     & 1.204 & 1.115 & 119.0 & 0 & 5.70 & Ref.~\citenum{hutter2003excited} \\
     \addlinespace[4mm]
     \multirow{2}{*}{${}^3A^{\prime\prime}$}& 1.3028 & 1.1057 & 113.19 & 42.77 & 2.686 & WEST-TDDFT \\
     & 1.305 & 1.108 & 113.7 & 43.2 & 2.67 & Ref.~\citenum{hutter2003excited} \\
     \addlinespace[1mm]
     \hline
   \end{tabular}
\end{table}

\subsection{NV$^-$ in Diamond}
\begin{table}
   \caption{Analytical and numerical gradients of the vertical excitation energy of the ${}^3E$ excited state of the NV$^-$ in diamond obtained using TDDFT with the LDA, PBE, and DDH functional. C29, C42, and C56 are three carbon atoms connected to the vacancy site. The gradients are provided in Ry bohr$^{-1}$.}
   \label{s-tbl:nvtriplet}
   \begin{tabular}{cccccccc}
     \hline
     \addlinespace[1mm]
     & Atom & \multicolumn{3}{c}{Analytical gradients} & \multicolumn{3}{c}{Numerical gradients} \\
     & & $X$ & $Y$ & $Z$ & $X$ & $Y$ & $Z$ \\
     \addlinespace[1mm]
     \hline
     \addlinespace[4mm]
     \multicolumn{8}{c}{State ${}^3E$} \\
     \addlinespace[4mm]
     & C29 & $-$0.022872 & \hphantom{$+$}0.006882 & \hphantom{$+$}0.020531 & $-$0.022847 & \hphantom{$+$}0.006893 & \hphantom{$+$}0.020535 \\
     LDA& C42 & $-$0.001743 & $-$0.046582 & \hphantom{$+$}0.001106 & $-$0.001742 & $-$0.046578 & \hphantom{$+$}0.001111 \\
     & C56 & $-$0.020777 & \hphantom{$+$}0.007246 & \hphantom{$+$}0.023305 & $-$0.020791 & \hphantom{$+$}0.007248 & \hphantom{$+$}0.023315 \\
     \addlinespace[4mm]
     & C29 & $-$0.028021 & \hphantom{$+$}0.010382 & \hphantom{$+$}0.015042 & $-$0.028024 & \hphantom{$+$}0.010373 & \hphantom{$+$}0.015040 \\
     PBE& C42 & \hphantom{$+$}0.000684 & $-$0.046292 & $-$0.001324 & \hphantom{$+$}0.000750 & $-$0.046226 & $-$0.001381 \\
     & C56 & $-$0.015312 & \hphantom{$+$}0.010708 & \hphantom{$+$}0.028343 & $-$0.015318 & \hphantom{$+$}0.010704 & \hphantom{$+$}0.028353 \\
     \addlinespace[4mm]
     & C29 & $-$0.038475 & \hphantom{$+$}0.017117 & \hphantom{$+$}0.008861 & $-$0.038528 & \hphantom{$+$}0.017107 & \hphantom{$+$}0.008816 \\
     DDH& C42 & \hphantom{$+$}0.001024 & $-$0.049392 & $-$0.001741 & \hphantom{$+$}0.001049 & $-$0.049284 & $-$0.001756 \\
     & C56 & $-$0.009310 & \hphantom{$+$}0.017345 & \hphantom{$+$}0.038643 & $-$0.009267 & \hphantom{$+$}0.017337 & \hphantom{$+$}0.038696 \\
     \addlinespace[1mm]
     \hline
   \end{tabular}
\end{table}
To verify our implementation of WEST-TDDFT, we compared TDDFT-TDA gradients of the vertical excitation energy obtained from the analytical expressions to those calculated from numerical differentiation for the triplet excited state ${}^3E$ (Table~\ref{s-tbl:nvtriplet}) and singlet excited states ${}^1E$ and ${}^1A_1$ (Table~\ref{s-tbl:nvsinglet}) of the NV$^-$ in diamond. The triplet excited state ${}^3E$ was studied using spin-conserving TDDFT while the singlet excited states ${}^1E$ and ${}^1A_1$ were studied using spin-flip TDDFT. The NV$^-$ center was simulated in a conventional $(2 \times 2 \times 2)$ supercell of diamond containing 63 atoms. The derivatives of the vertical excitation energy were calculated with respect to the atomic coordinates of carbon atoms C29, C42, and C56, which are the three carbon atoms connected to the vacancy site. The gradients of the ${}^3E$ state were computed at the optimized geometry of the ${}^3E$ state, and the gradients of the ${}^1E$ and ${}^1A_1$ states were computed at the optimized geometry of the ${}^1E$ state; both geometries were obtained using TDDFT with the PBE functional. The numerical gradients were computed using a finite difference approach with a displacement of 0.007136 \AA. The Quantum ESPRESSO code~\cite{giannozzi2020qe,carnimeo2023quantum} was employed for the ground-state DFT calculations, with the LDA~\cite{perdew1981self}, PBE~\cite{perdew1996generalized} or DDH~\cite{skone2014ddh,skone2016ddh} functional, the SG15 ONCV pseudopotentials~\cite{hamann2013optimized,schlipf2015optimization}, and a plane-wave energy cutoff of 60 Ry. To prevent the numerical instabilities in the spin-flip TDDFT calculations, we set values of $f_{\text{xc}}^{\text{loc,SF}}$ to zero on a grid point $\mathbf{r}_0$ if the absolute value of the spin density $\rho_{\uparrow}^0 (\mathbf{r}_0) - \rho_{\downarrow}^0 (\mathbf{r}_0)$ on this point is smaller than $1\times10^{-3}$. As shown in Table~\ref{s-tbl:nvtriplet} and Table~\ref{s-tbl:nvsinglet}, the analytical and numerical gradients agree very well with each other.
\begin{table}
   \caption{Analytical and numerical gradients of the vertical excitation energy of the ${}^1E$ and ${}^1A_1$ excited states of the NV$^-$ in diamond obtained using TDDFT with the LDA, PBE, and the DDH functional. C29, C42, and C56 are three carbon atoms connected to the vacancy site. The gradients are provided in Ry bohr$^{-1}$.}
   \label{s-tbl:nvsinglet}
   \begin{tabular}{cccccccc}
     \hline
     \addlinespace[1mm]
     & Atom & \multicolumn{3}{c}{Analytical gradients}  & \multicolumn{3}{c}{Numerical gradients}  \\
     & & $X$ & $Y$ & $Z$ & $X$ & $Y$ & $Z$ \\
     \addlinespace[1mm]
     \hline
     \addlinespace[3mm]
     \multicolumn{8}{c}{State ${}^1E$} \\
     \addlinespace[3mm]
     & C29 & \hphantom{$+$}0.002100 & \hphantom{$+$}0.002938 & \hphantom{$+$}0.002176 & \hphantom{$+$}0.002100 & \hphantom{$+$}0.002946 & \hphantom{$+$}0.002175 \\
     LDA& C42 & \hphantom{$+$}0.008544 & $-$0.014794 & $-$0.009486 & \hphantom{$+$}0.008533 & $-$0.014795 & $-$0.009471 \\
     & C56 & $-$0.002373 & \hphantom{$+$}0.003749 & $-$0.001157 & $-$0.002368 & \hphantom{$+$}0.003764 & $-$0.001141 \\
     \addlinespace[4mm]
     & C29 & \hphantom{$+$}0.004743 & \hphantom{$+$}0.002861 & \hphantom{$+$}0.006857 & \hphantom{$+$}0.004695 & \hphantom{$+$}0.002894 & \hphantom{$+$}0.006824 \\
     PBE& C42 & \hphantom{$+$}0.014139 & $-$0.024152 & $-$0.014744 & \hphantom{$+$}0.014091 & $-$0.023985 & $-$0.014677 \\
     & C56 & $-$0.006685 & \hphantom{$+$}0.003664 & $-$0.003715 & $-$0.006630 & \hphantom{$+$}0.003676 & $-$0.003677 \\
     \addlinespace[4mm]
     & C29 & \hphantom{$+$}0.008305 & \hphantom{$+$}0.003743 & \hphantom{$+$}0.013414 & \hphantom{$+$}0.008263 & \hphantom{$+$}0.003796 & \hphantom{$+$}0.013369 \\
     DDH& C42 & \hphantom{$+$}0.020372 & $-$0.035543 & $-$0.020950 & \hphantom{$+$}0.020291 & $-$0.035246 & $-$0.020892 \\
     & C56 & $-$0.012993 & \hphantom{$+$}0.004753 & $-$0.006982 & $-$0.012924 & \hphantom{$+$}0.004808 & $-$0.006935 \\
     \addlinespace[3mm]
     \multicolumn{8}{c}{State ${}^1A_1$} \\
     \addlinespace[3mm]
     & C29 & $-$0.009928 & \hphantom{$+$}0.000453 & $-$0.017409 & $-$0.009918 & \hphantom{$+$}0.000466 & $-$0.017420 \\
     LDA & C42 & \hphantom{$+$}0.008165 & $-$0.001599 & $-$0.007678 & \hphantom{$+$}0.008194 & $-$0.001603 & $-$0.007695 \\
     & C56 & \hphantom{$+$}0.017676 & \hphantom{$+$}0.000242 & \hphantom{$+$}0.009756 & \hphantom{$+$}0.017684 & \hphantom{$+$}0.000260 & \hphantom{$+$}0.009758 \\
     \addlinespace[4mm]
     & C29 & $-$0.008408 & \hphantom{$+$}0.005775 & $-$0.014338 & $-$0.008413 & \hphantom{$+$}0.005788 & $-$0.014333 \\
     PBE & C42 & \hphantom{$+$}0.009275 & $-$0.003203 & $-$0.009050 & \hphantom{$+$}0.009423 & $-$0.003439 & $-$0.009179 \\
     & C56 & \hphantom{$+$}0.014471 & \hphantom{$+$}0.005658 & \hphantom{$+$}0.008310 & \hphantom{$+$}0.014469 & \hphantom{$+$}0.005670 & \hphantom{$+$}0.008314 \\
     \addlinespace[4mm]
     & C29 & $-$0.005726 & \hphantom{$+$}0.009158 & $-$0.012240 & $-$0.005730 & \hphantom{$+$}0.009176 & $-$0.012243 \\
     DDH & C42 & \hphantom{$+$}0.010072 & $-$0.003079 & $-$0.009980 & \hphantom{$+$}0.010186 & $-$0.003292 & $-$0.010082 \\
     & C56 & \hphantom{$+$}0.012301 & \hphantom{$+$}0.009101 & \hphantom{$+$}0.005671 & \hphantom{$+$}0.012300 & \hphantom{$+$}0.009121 & \hphantom{$+$}0.005673 \\
     \addlinespace[1mm]
     \hline
   \end{tabular}
\end{table}
\newpage

\section{Details on the Scalability of WEST-TDDFT}

We present an assessment of the performance and scalability of the WEST-TDDFT code on both CPU and GPU nodes, using the Perlmutter supercomputer at the National Energy Research Scientific Computing Center (NERSC). The CPU nodes of Perlmutter are equipped with two AMD EPYC Milan CPUs, with a theoretical peak double-precision (FP64) performance of 5.0 TFLOPS (tera floating-point operations per second) per node. The GPU nodes of Perlmutter are equipped with one AMD EPYC Milan CPU and four NVIDIA A100 GPUs, with a theoretical peak FP64 performance of 39.0 TFLOPS per node, or 78.0 TFLOPS if the acceleration provided by the tensor cores is used. The latter are specialized computing units optimized for matrix multiplications. Comparing the workflows of full-frequency $G_0W_0$ and TDDFT implemented in WEST, we observe a higher utilization rate of the tensor core in the former because its implementation involved more matrix multiplication operations.

To benchmark the implementation of WEST-TDDFT, we considered the first excited state of the NV$^-$ in diamond, whose ground-state DFT calculations were performed using the Quantum ESPRESSO~\cite{giannozzi2020qe,carnimeo2023quantum} code (version 7.2), the SG15 optimized norm-conserving Vanderbilt (ONCV) pseudopotentials~\cite{hamann2013optimized,schlipf2015optimization}, and the DDH functional~\cite{skone2014ddh,skone2016ddh}. A kinetic energy cutoff of 60 Ry was used for the plane-wave basis set. The Brillouin zone was sampled at the $\Gamma$-point. In the TDDFT calculations, we used the numerical approximations discussed in Section~3 of the main text, namely the ACE method with $N_{\text{ACE}} = 4 N_{\text{occ}}$, the Wannier localization with $S_{\text{thr}} = 10^{-3}$, and the inexact Krylov subspace approach with $\lambda_{\text{thr}} = 10^{-4}$.

\begin{figure}
    \centering
    \includegraphics[width=16cm]{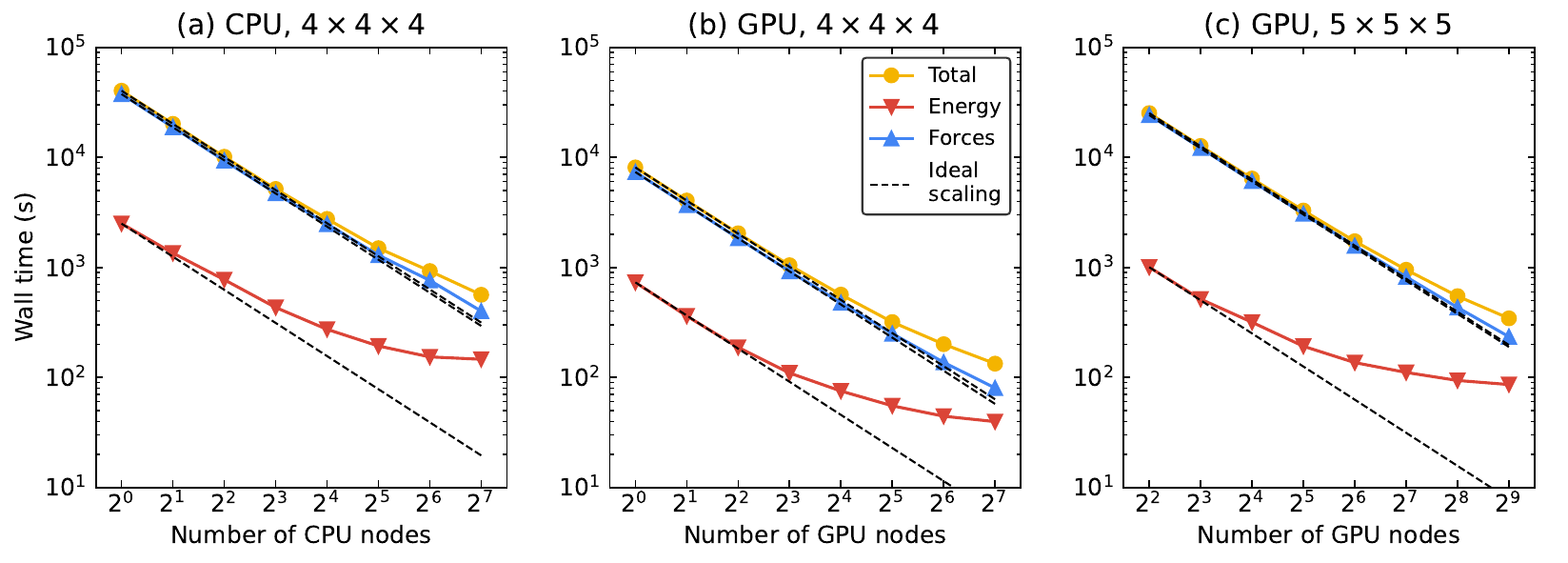}
    \caption{Strong scaling of the WEST-TDDFT code on CPU and GPU nodes of NERSC/Perlmutter supercomputing architecture. The scaling tests were carried out on the electronic properties of the first excited state of the NV$^-$ center in diamond, simulated using the DDH functional and 60 Ry of kinetic energy cutoff. (a) NV$^-$ in a conventional $(4 \times 4 \times 4)$ supercell of diamond. The calculations were carried out on CPU nodes. (b) NV$^-$ in a conventional $(4 \times 4 \times 4)$ supercell of diamond. The calculations were carried out on GPU nodes. (c) NV$^-$ in a conventional $(5 \times 5 \times 5)$ supercell of diamond. The calculations were carried out on GPU nodes. The red downward triangles and blue upward triangles represent the time measured to compute the vertical excited state energy and forces, respectively. The yellow circles represent the total wall clock time, including the time spent on I/O operations. The black dashed lines indicate ideal strong scaling.}
    \label{s-fig:scaling_nv}
\end{figure}

The performance of the CPU version of WEST-TDDFT is presented in Figure~\ref{s-fig:scaling_nv} (a) for the NV$^-$ center in a conventional $(4 \times 4 \times 4)$ supercell of diamond containing 511 atoms. The figure displays the total wall clock time including time spent on I/O operations, and breaks it down into the time required to compute energy or forces. For the part of the code that computes forces (see blue upward triangles) we observe very good strong scaling up to 128 nodes, i.e., a scaling that closely aligns with the ideal one (indicated by the black dashed line). For the part of the code that computes energies (see red downward triangles) we observe good scaling up to 32 nodes, after which the parallel efficiency drops because the computational time becomes small and comparable to the overheads caused by internode MPI communications and I/O operations. Nonetheless, the overall strong scaling of WEST-TDDFT (corresponding to the time required to compute both energies and forces, and represented by orange circles), remains close to ideal, as the computation of forces is approximately one order of magnitude more expensive than the computation of energy. 

In Figure~\ref{s-fig:scaling_nv} (b), we examine the performance of the GPU version of WEST-TDDFT for the same test system of 511 atoms. Again, the code exhibits nearly ideal scaling up to 128 GPU nodes (512 GPUs), primarily due to the highly efficient implementation of the forces. Similar to the CPU case, the scaling of the energy part, which has a lower computation-to-communication ratio, is inferior to that of the forces. Comparing Figure~\ref{s-fig:scaling_nv} (a) and (b), we observe that the GPU version of WEST-TDDFT achieves a 4.8$\times$ overall speedup over its CPU counterpart on the same number of nodes, with the computation of the energy and the forces accelerated by 3.7$\times$ and 5.2$\times$, respectively. The higher speedup achieved in computing the forces is due to the larger amount of computations involved in this part. We note that the attained speedup falls short of the theoretical speedup estimated by considering the ratio between the theoretical peak performance of the GPU and CPU nodes. This discrepancy is primarily attributed to (i) the communication overhead when summing over the bands, which becomes necessary each time the operators on the left hand side of Eq.~(1) are applied, and (ii) the overhead associated with reading the ground state wavefunctions from the file system and writing the excited state wavefunctions to the file system. The scalability of the GPU version of WEST-TDDFT is further demonstrated in Figure~\ref{s-fig:scaling_nv} (c) for the NV$^-$ center in a $(5 \times 5 \times 5)$ supercell of diamond containing 999 atoms. Remarkably, WEST-TDDFT scales to 512 GPU nodes (2048 GPUs), with the forces part of the code exhibiting near-perfect strong scaling.

\newpage

\section{Excited-State Geometry Relaxation of the SiV$^0$ in Diamond}
The geometry relaxation was carried out in the triplet excited state for SiV$^0$ in diamond using TDDFT-TDA with the PBE and DDH functional. We used $(3 \times 3 \times 3)$, $(4 \times 4 \times 4)$, and $(5 \times 5 \times 5)$ conventional supercells of diamond that contain 215, 511, and 999 atoms, respectively. Results on relaxed geometries are summarized in Table~\ref{s-tbl:relaxsiv0}. The zero-phonon line ($E_{\text{ZPL}}$) and the Franck-Condon shift in electronic excited states ($E_{\text{FC, ES}}$) computed using TDDFT with the DDH functional agree well with the $\Delta$SCF results reported in Ref.~\citenum{thiering2019eg}.

Franck-Condon shifts in electronic excited states $E_{\text{FC, ES}}$ and electronic ground states $E_{\text{FC, GS}}$, and mass-weighted atomic displacements $\Delta Q$ computed using TDDFT with the PBE functional exhibit non-negligible finite-size effects, while those obtained with the DDH functional do not. This discrepancy stems from different degrees of localization of the excited state orbitals obtained with the PBE and DDH functional. The exciton is more delocalized with the PBE functional, and its radius increases as the cell size increases, and as a consequence, the magnitude of geometry displacements decreases. The exciton obtained with the DDH functional is almost fully localized within the $(3 \times 3 \times 3)$ supercell, and increasing the cell size brings a negligible change to the magnitude of geometry displacements.
\begin{table}
   \caption{Computed zero-phonon line energies $E_{\text{ZPL}}$ (eV), Franck-Condon shifts (eV) in electronic excited states $E_{\text{FC, ES}}$ and electronic ground states $E_{\text{FC, GS}}$, mass-weighted atomic displacements $\Delta Q$ (amu$^{1/2}$ \AA ), and Si$-$C bond lengths (\AA) for SiV$^0$ in diamond.}
   \label{s-tbl:relaxsiv0}
   \begin{tabular}{llllllll}
     \hline
     Method & Cell size & $E_{\text{ZPL}}$ & $E_{\text{FC, GS}}$ & $E_{\text{FC, ES}}$ & $\Delta Q$ & $d$ (Si$-$C) \\
     \hline
     TDDFT (PBE) & $(3 \times 3 \times 3)$ & 1.501 & 0.236 & 0.265 & 0.510 & 1.965, 2.051 \\
     TDDFT (PBE) & $(4 \times 4 \times 4)$ & 1.299 & 0.145 & 0.142 & 0.412 & 1.966, 2.029 \\
     TDDFT (PBE) & $(5 \times 5 \times 5)$ & 1.124 & 0.079 & 0.086 & 0.289 & 1.970, 2.005 \\
     TDDFT (DDH) & $(3 \times 3 \times 3)$ & 1.518 & 0.314 & 0.338 & 0.581 & 1.952, 2.052 \\
     TDDFT (DDH) & $(4 \times 4 \times 4)$ & 1.403 & 0.298 & 0.294 & 0.590 & 1.953, 2.050 \\
     $\Delta$SCF (HSE)~\cite{thiering2019eg} & $(4 \times 4 \times 4)$ & 1.34 & & 0.258 \\
     Expt.~\cite{d2011optical} & & 1.31 \\
     \hline
   \end{tabular}
 \end{table}
\newpage

\section{Extrapolation of Vertical Excitation Energies of the SiV0 in Diamond}
\subsection{Defect Excitation}
\begin{figure}
    \centering
    \includegraphics[width=13cm]{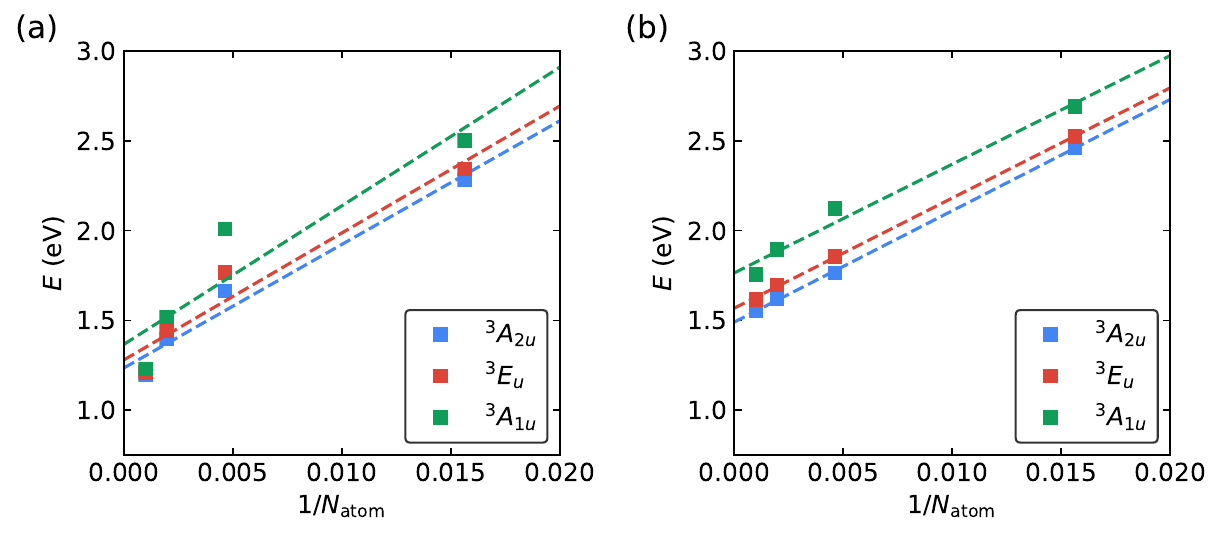}
    \caption{Vertical excitation energies (VEEs) of the triplet excited states ${}^3A_{2u}$ (blue), ${}^3E_{u}$ (red) and ${}^3A_{1u}$ (green) for the SiV$^0$ in diamond as a function of the number of atoms $N_\text{atoms}$ in the supercell. The VEEs were obtained using spin-conserving TDDFT calculations with (a) the PBE functional and (b) the DDH functional. Dashed lines show linear extrapolations of VEEs as a function of $1/N_{\text{atoms}}$.}
    \label{s-fig:siv0triplet}
\end{figure}

\begin{figure}
    \centering
    \includegraphics[width=13cm]{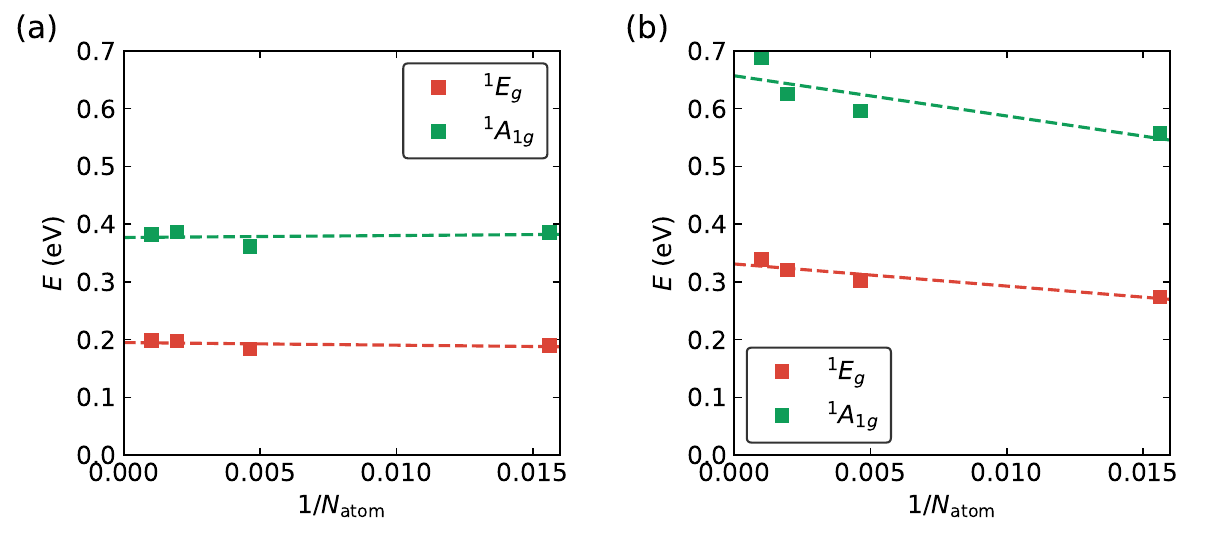}
    \caption{Vertical excitation energies (VEEs) of the singlet excited states ${}^1E_{g}$ (red) and ${}^1A_{1g}$ (green) for the SiV$^0$ in diamond as a function of the number of atoms $N_\text{atoms}$ in the supercell. The VEEs were obtained using spin-flip TDDFT calculations with (a) the PBE functional and (b) the DDH functional. Dashed lines show linear extrapolations of VEEs as a function of $1/N_{\text{atoms}}$.}
    \label{s-fig:siv0singlet}
\end{figure}

The vertical excitation energies (VEEs) for excited states based on transitions among defect orbitals are summarized in Figure~\ref{s-fig:siv0triplet} and Figure~\ref{s-fig:siv0singlet} for triplet and singlet excited states, respectively. The VEEs were computed using TDDFT with the PBE and DDH functional at the optimized ground state atomic geometry and were extrapolated to the dilute limit using linear functions of $1/N_{\text{atoms}}$, where $N_{\text{atoms}}$ is the number of atoms in the supercell. The conventional $(2\times2\times2)$, $(3\times3\times3)$, $(4\times4\times4)$, and $(5\times5\times5)$ diamond supercells were used in the calculations of VEEs, which contains 63, 215, 511, and 999 atoms, respectively.


The VEEs of singlet excited states ${}^1E_g$ and ${}^1A_{1g}$ depend weakly on $N_{\text{atoms}}$ since the composition of their respective many-body wavefunctions mainly involve transitions from $e_{gx}(e_{gy})$ orbitals to $\overline{e}_{gx}(\overline{e}_{gy})$ orbitals, which are all within the band gap of diamond. In contrast, the VEEs of triplet excited states ${}^3A_{2u}$, ${}^3E_{u}$ and ${}^3A_{1u}$ exhibit a linear dependence on $1/N_{\text{atoms}}$, which stems from the dipole-dipole interaction between the localized excitons in periodic images. The triplet excited states computed with the DDH functional show a clearer linear relationship for VEEs than those with the PBE functional. This may be due to more localized excitons computed with the DDH functional, which are better described by the model accounting for the dipole-dipole interaction in periodic images.

\subsection{Bound Exciton}
\label{s-sec:siv0boundexcitonfitting}
To extrapolate the VEE of the bound exciton state, which is based on the transition from the valence band maximum (VBM) to the $\overline{e}_{gs}/\overline{e}_{gy}$ defect orbitals, we used the fitting function
\begin{equation}
    \label{s-eq:siv0boundexcitonfitting}
    E_{\text{VE}}(L) = E_{\text{VE}}(L = \infty) - \frac{A}{L} \exp \left(-\frac{L}{D}\right) + \frac{B}{L^3}.
\end{equation}
where $L \propto N_{\text{atoms}}^{1/3}$ is the length of the cubic supercell, $A$ and $B$ are fitting parameters determined by the strength of the electron-hole interaction and the dipole-dipole interaction, respectively. $D$ is a screening parameter related to the fact that the electron-hole interaction does not depend on $L$ if $L$ is much larger than the radius of the bound exciton.

\begin{figure}
    \centering
    \includegraphics[width=16cm]{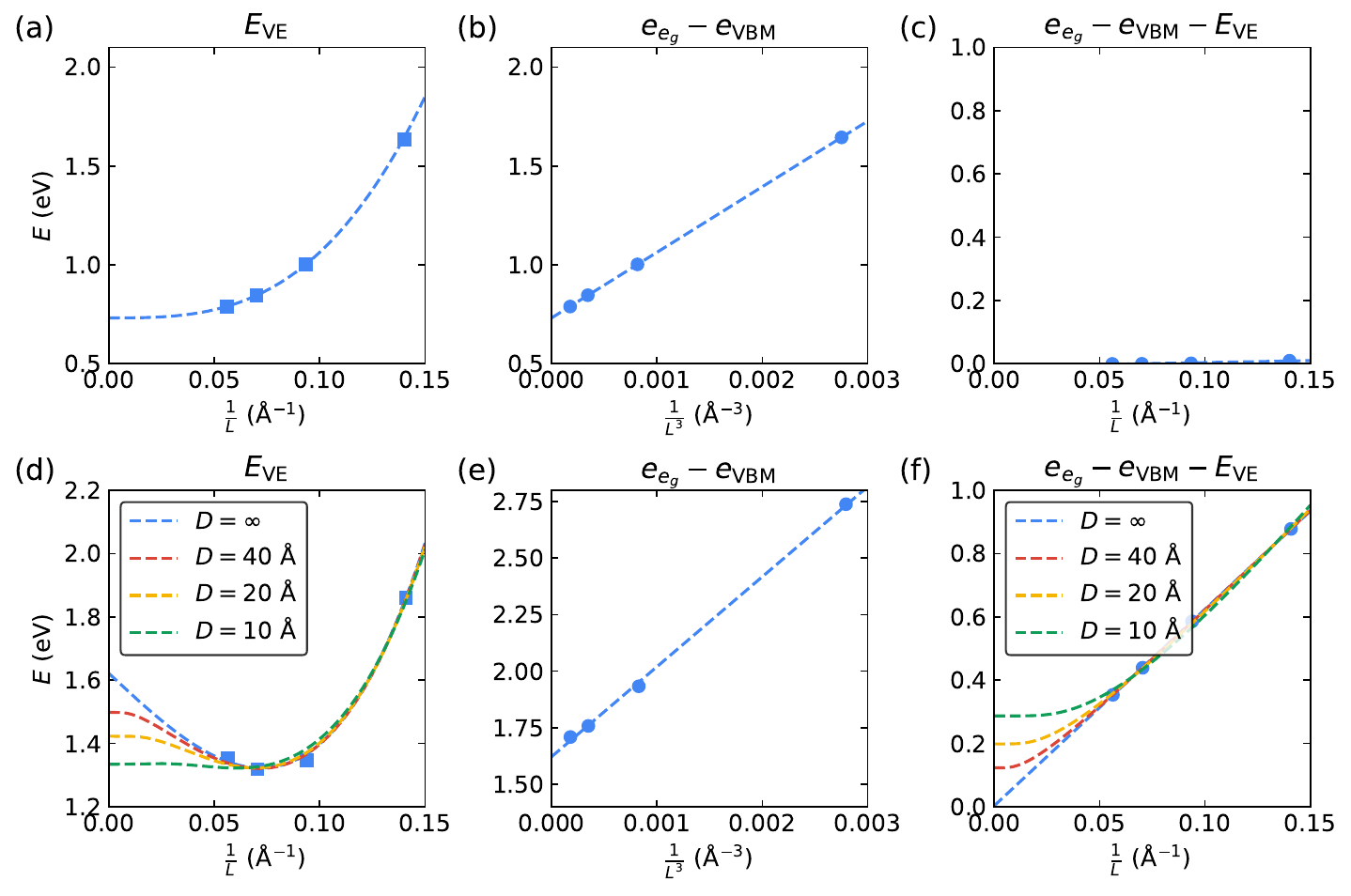}
    \caption{Extrapolation of the vertical excitation energy $E_{\text{VE}}$ for the bound exciton of SiV$^0$ in diamond to the dilute limit. $E_{\text{VE}}$ obtained from TDDFT with the PBE functional and the DDH functional are displayed in (a) and (d) as a function of $\frac{1}{L}$. The values of $E_{\text{VE}}$ are fitted using Eq.~\eqref{s-eq:siv0boundexcitonfitting}, and the fitting lines are shown as dashed lines. The energy difference of the defect orbital $e_g$ and the valence band maximum (VBM), $e_{e_g} - e_{\text{VBM}}$, obtained using the PBE functional and the DDH functional are displayed in (b) and (e) as a function of $\frac{1}{L^3}$. The exciton binding energy, $e_{e_g} - e_{\text{VBM}} - E_{\text{VE}}$, obtained using the PBE functional and the DDH functional are displayed in (c) and (f) as a function of $\frac{1}{L}$. Details on the fitting process can be found in Section~\ref{s-sec:siv0boundexcitonfitting}.}
    \label{s-fig:siv0boundexciton}
\end{figure}

To obtain the fitting parameter $B$, we decompose the VEE into two terms. The first term is $e_{e_g} - e_{\text{VBM}}$, which can be approximately viewed as the fundamental gap for the bound exciton. The second term is $e_{e_g} - e_{\text{VBM}} - E_{\text{VE}}$, which can be approximately viewed as the exciton binding energy, i.e. the difference between the fundamental gap and the optical gap. Figure~\ref{s-fig:siv0boundexciton} shows the VEE, the fundamental gap, and the exciton binding energy as functions of supercell size. The fundamental gap $e_{e_g} - e_{\text{VBM}}$ shows a linear dependence on $1/L^3$, and the slope yields $B = 331.8$ eV \AA$^3$ for the PBE functional and $B = 398.3$ eV \AA$^3$ for the DDH functional.

\newpage

To obtain the fitting parameter $D$, which is the effective screening length for the bound exciton, we adopted the model proposed in Ref.~\citenum{zhang2020optically}. The screening length is expressed as
\begin{equation}
\label{s-eq:screeninglength}
    D = \alpha \dfrac{m_e \varepsilon_\infty}{m^{\ast}} D_\text{H},
\end{equation}
where $D_{\text{H}} = 1.9$ \AA\ is the screening length of the hydrogen atom obtained from DFT calculations~\cite{zhang2020optically}, $m^{\ast} = 0.7 m_e \ (2.12 m_e)$ is the effective mass of light (heavy) hole in VBM, $\varepsilon_{\infty} = 5.7$ is the dielectric constant of diamond. $\alpha = 1.87$ is derived from the comparison with experiments to account for the fact that the hole orbital is expelled from the central region of the defect as the $e_g$ orbital of the electron already occupies this region~\cite{zhang2020optically}. With these parameters, the screening length is estimated as $D \in [10, 40]$ \AA.

To obtain the fitting parameter $A$, we fitted the exciton binding energy $e_{e_g} - e_{\text{VBM}} - E_{\text{VE}}$ as $\frac{A}{L}\exp\left(-\frac{L}{D}\right)$. With $D = \infty$, the exciton binding energy was fitted as a function versus against $1/L$, and the slope yields $A = 0.12$ eV \AA\ for the PBE functional and $A = 6.21$ eV \AA\ for the DDH functional. The intercept of the linear function is almost zero, pointing to the fact that the exciton binding energy is zero for an infinite large exciton whose screening length is $D = \infty$. It is worth noting that the exciton binding energy predicted by TDDFT with the PBE functional is approximately zero because the Coulomb interaction between the electron and the hole of the exciton is absent with the PBE functional, which points to the necessity of using the DDH functional in TDDFT calculations. By setting $D = \infty$ in Eq.~\eqref{s-eq:siv0boundexcitonfitting}, we obtained $E_{\text{VE}}(L = \infty) = 1.621$ eV, which corresponds to the charge transition level of the $\text{SiV}^- + h^+(\text{at VBM}) \to \text{SiV}^0$ process. With $D \in [10, 40]$ \AA, we obtained $A \in [6.41, 8.65]$ eV \AA, which yields $E_{\text{VE}} (L = \infty) \in [1.33, 1.50]$ eV.
\newpage

\section{Extrapolation of Vertical Excitation Energies of the V$_{\text{O}}^0$ in MgO}
\subsection{Defect Excitation}
\begin{figure}
    \centering
    \includegraphics[width=13cm]{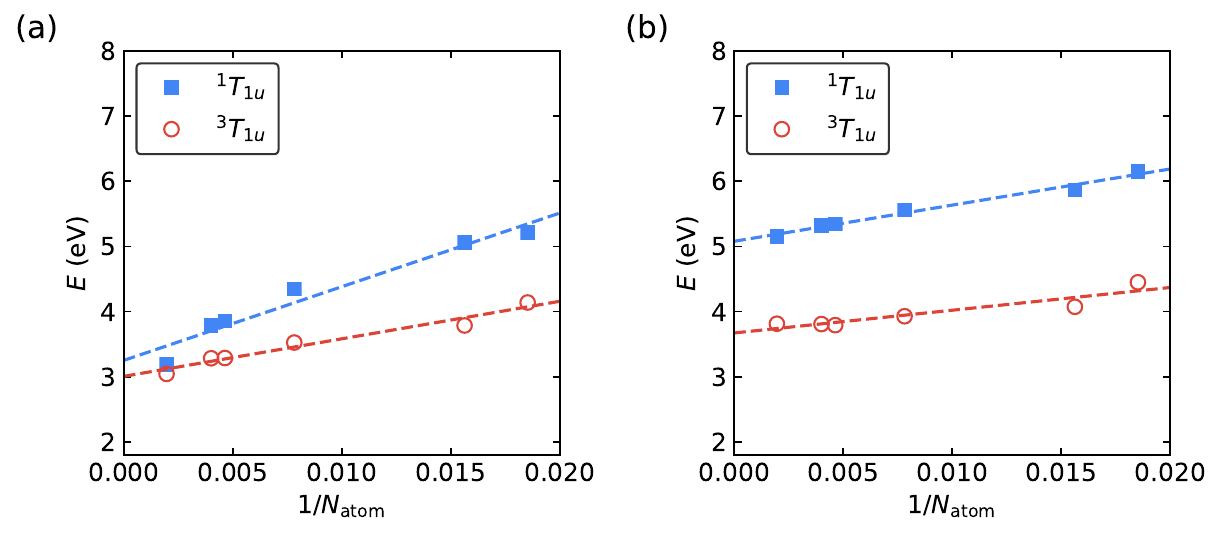}
    \caption{Vertical excitation energies (VEEs) of the singlet and triplet excited states ${}^1T_{1u}$ (blue squares) and ${}^3T_{1u}$ (red circles) for the V$_{\text{O}}^0$ in MgO as a function of the number of atoms, $N_\text{atoms}$, in the supercell. The VEEs were obtained using TDDFT calculations with (a) the PBE functional and (b) the DDH functional. Dashed lines show linear extrapolations of VEEs based on $1/N_{\text{atoms}}$.}
    \label{s-fig:vost}
\end{figure}

The vertical excitation energies (VEEs) for excited states based on transitions among defect orbitals are summarized in Figure~\ref{s-fig:vost} for triplet and singlet excited states. The VEEs were computed using TDDFT with the PBE functional and the DDH functional at the optimized ground state atomic geometry and were extrapolated to the dilute limit as linear functions of $1/N_{\text{atoms}}$, where $N_{\text{atoms}}$ is the number of atoms in the supercell. The primitive $(3\times3\times3)$, $(4\times4\times4)$ and $(5\times5\times5)$ MgO supercells containing 53, 127 and 249 atoms, and the conventional $(2\times2\times2)$, $(3\times3\times3)$ and $(4\times4\times4)$ MgO supercells containing 63, 215 and 511 atoms were used in the calculations of VEEs.

The VEEs of singlet and triplet excited states ${}^1T_{1u}$ and ${}^3T_{1u}$ exhibit a linear dependence on $1/N_{\text{atoms}}$, which stems from the dipole-dipole interaction between the localized excitons in periodic images.

\subsection{Bound Exciton}
The VEE of the bound exciton state of V$_{\text{O}}^0$ in MgO, which is based on the transition from the $a_{1g}$ defect orbital to the conduction band minimum (CBM), is extrapolated using the same strategy as the one used for the SiV$^0$ in diamond, as summarized in Section~\ref{s-sec:siv0boundexcitonfitting}. For conventional supercells, $L$ is simply the length of the cubic supercell. For primitive supercells, $L$ is defined as $L = \left(N_{\text{atoms}} / 8\right)^{1/3} L_0$, where $L_0 = 4.26$ \AA\ is the lattice constant of MgO obtained with the DDH functional. $A = 10.60$ eV \AA\ was obtained for the DDH functional by fitting the exciton binding energy $e_{\text{CBM}} - e_{a_{1g}} - E_{\text{VE}}$ as a linear function of $1 / L$ with the intercept fixed as zero. $B = 462.2$ eV \AA$^3$ was obtained for the DDH functional by fitting $e_{\text{CBM}} - e_{a_{1g}}$ as a linear function of $1 / L^3$. The screening length is estimated as $D \in [21, 42]$ \AA\ using Eq.~\eqref{s-eq:screeninglength}, with $\varepsilon_{\infty} = 2.96$~\cite{skone2014ddh}, $m^{\ast} = 0.4 m_e$~\cite{schleife2012ab}, and an estimated $\alpha \in [1.5, 3]$. With these parameters, the vertical excitation energy of the bound exciton state is estimated to be $[4.41, 4.58]$ eV for the DDH functional.
\label{s-sec:mgoboundexcitonfitting}
\begin{figure}
    \centering
    \includegraphics[width=16cm]{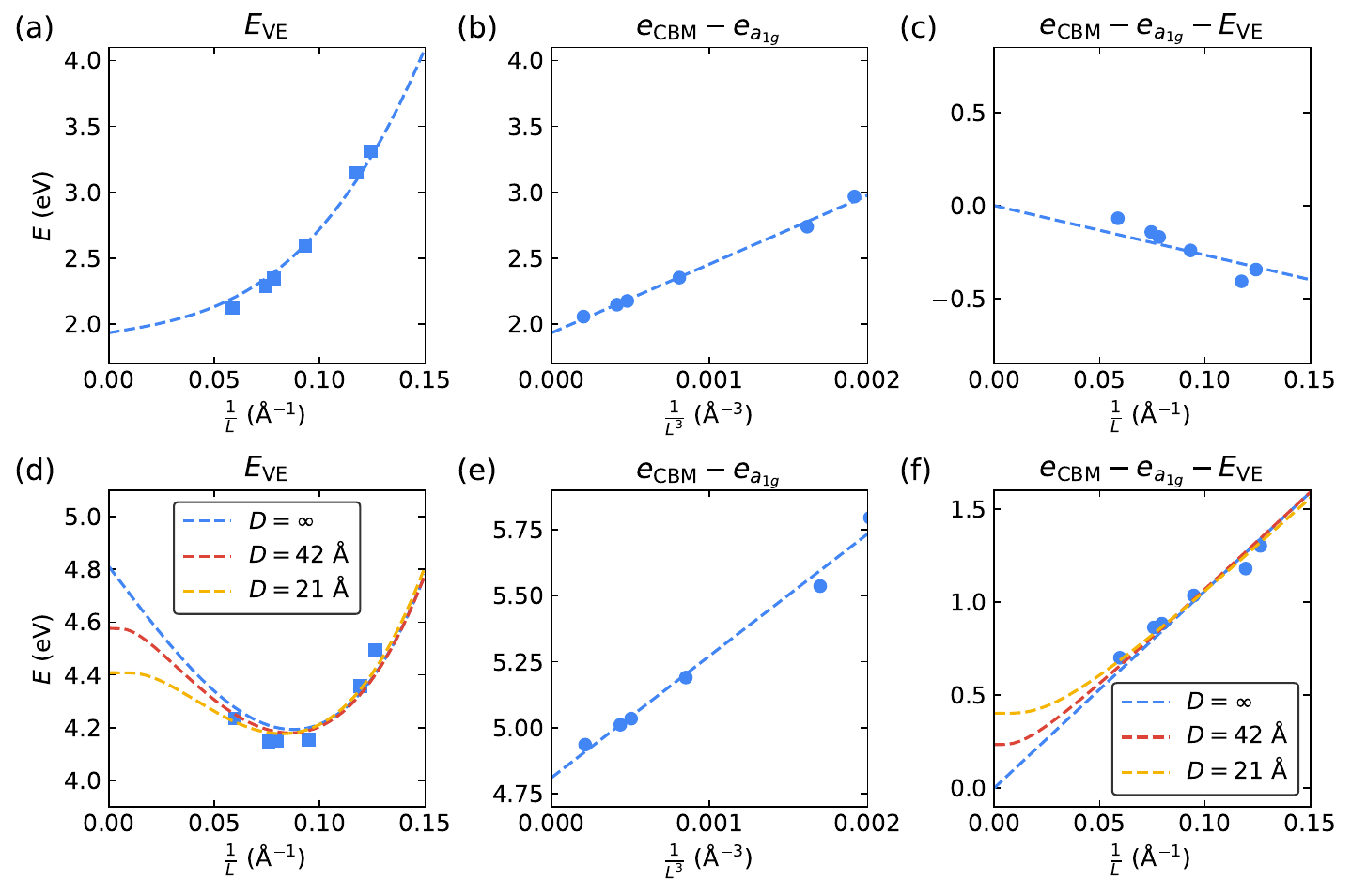}
    \caption{Extrapolation of the vertical excitation energy $E_{\text{VE}}$ for the bound exciton of V$_{\text{O}}^0$ in MgO to the dilute limit. $E_{\text{VE}}$ obtained from TDDFT with the PBE functional and the DDH functional are displayed in (a) and (d) as a function of $\frac{1}{L}$. The values of $E_{\text{VE}}$ are fitted using Eq.~\eqref{s-eq:siv0boundexcitonfitting}, and the fitting lines are shown as dashed lines. The energy difference of the conduction band minimum (CBM) and the defect orbital $a_{1g}$, $e_{\text{CBM}} - e_{a_{1g}}$, obtained using the PBE functional and the DDH functional are displayed in (b) and (e) as a function of $\frac{1}{L^3}$. The exciton binding energy, $e_{\text{CBM}} - e_{a_{1g}} - E_{\text{VE}}$, obtained using the PBE functional and the DDH functional are displayed in (c) and (f) as a function of $\frac{1}{L}$. Details on the fitting process can be found in Section~\ref{s-sec:mgoboundexcitonfitting}.}
    \label{s-fig:voboundexciton}
\end{figure}

\begin{table}
   \caption{Computed absorption energy $E_{\text{Abs}}$ (eV), emission energy $E_{\text{Emi}}$, Franck-Condon shifts (eV) in electronic excited states $E_{\text{FC, ES}}$ and electronic ground states $E_{\text{FC, GS}}$, and mass-weighted atomic displacements $\Delta Q$ (amu$^{1/2}$ \AA ) for V$_{\text{O}}^0$ in MgO. Absorption energies were computed using TDDFT with the PBE functional and extrapolated to the dilute limit. FC shifts and displacements were computed using TDDFT with the PBE functional and a conventional $(3 \times 3 \times 3)$ supercell with 215 atoms.}
   \label{s-tbl:relaxvo0pbe}
   \begin{tabular}{lllllll}
     \hline
     States & $E_{\text{Abs}}$ & $E_{\text{Emi}}$ & $E_{\text{FC, GS}}$ & $E_{\text{FC, ES}}$ & $\Delta Q$ \\
     \hline
     ${}^1T_{1u}$ & 3.254 & 2.307 & 0.410 & 0.537 & 1.694 \\
     ${}^3T_{1u}$ & 3.008 & 2.387 & 0.263 & 0.358 & 2.224 \\
     Bound Exciton & 1.934 & 1.444 & 0.182 & 0.308 & 0.659 \\
     \hline
   \end{tabular}
\end{table}
\newpage


\bibliography{SI_v1.bib}